\documentclass[journal]{IEEEtran}
\ifCLASSINFOpdf
\else
\fi

\usepackage{graphics}
\usepackage{epsfig}
\usepackage{mathptmx}
\usepackage{cite}
\usepackage{amsmath}
\usepackage{amssymb}
\usepackage{algorithm}
\usepackage{algorithmic}
\usepackage{array}
\usepackage{stfloats}
\usepackage{url}
\usepackage{multirow}
\usepackage{bbm}

\renewcommand{\baselinestretch}{0.97}

\begin{document}
%

\title{Assessing the Impacts of Nonideal Communications on Distributed Optimal Power Flow Algorithms}

%

\author{Mohannad~Alkhraijah,~\IEEEmembership{Student Member,~IEEE,}
        Carlos~Menendez,~\IEEEmembership{Student Member,~IEEE,}
        and~Daniel~K~Molzahn,~\IEEEmembership{Senior~Member,~IEEE}
}

\maketitle

\begin{abstract}
Power system operators are increasingly looking toward distributed optimization to address various challenges facing electric power systems. To assess their capabilities in environments with nonideal communications, this paper investigates the impacts of data quality on the performance of distributed optimization algorithms. Specifically, this paper compares the performance of the Alternating Direction Method of Multipliers (ADMM), Analytical Target Cascading (ATC), and Auxiliary Problem Principle (APP) algorithms in the context of DC Optimal Power Flow (DC~OPF) problems. Using several test systems, this paper characterizes the performance of these algorithms in terms of their convergence rates and solution quality under three data quality nonidealities: (1)~additive Gaussian noise, (2)~bad data (large error), and (3)~intermittent communication failure.
\end{abstract}


%
\IEEEpeerreviewmaketitle

\section{Introduction}
%
%
%
%
\IEEEPARstart{T}{raditionally}, power system operations are primarily conducted centrally, where the tasks of modeling the system's components, solving an optimization problem, and dispatching setpoints are performed by the system operator. However, as power systems move toward a more decentralized paradigm with many locally controlled microgrids, scaling these tasks becomes challenging when using centralized methods. Further, modeling connected systems is difficult, especially for microgrids in a distribution network. 

Distributed algorithms address these challenges by allowing interconnected systems with local controllers to cooperatively solve large optimization problems. This increases flexibility as each microgrid solves a local subset of the original optimization problem. Distributed algorithms thus have various potential advantages, such as allowing parallel computations, reducing the requisite communication infrastructure, and possibly maintaining the subsystems' autonomy and data privacy. 

These advantages motivate solving Optimal Power Flow (OPF) problems in a distributed fashion. OPF problems seek optimal setpoints for a power system. The decision variables are typically the generators' power outputs and the voltage phasors. Several distributed optimization algorithms have been proposed for solving OPF problems~\cite{molzahn2017survey}. The Alternating Direction Method of Multipliers (ADMM) applies Augmented Lagrangian Relaxation to decompose the centralized problems into smaller problems which permit a distributed implementation~\cite{boyd2011distributed}. Augmented Lagrangian Relaxation is also used in other distributed OPF solution algorithms, such as Analytical Target Cascading (ATC)~\cite{8240623}, Auxiliary Problem Principle (APP)~\cite{kim1997coarse}, and Dual Decomposition. Other distributed algorithms directly consider the Karush-Kuhn-Tucker (KKT) optimality conditions for OPF problems. These include the Optimality Condition Decomposition and Consensus+Innovation algorithms. Numerical comparisons among these distributed OPF algorithms are presented in~\cite{kim2000comparison} and~\cite{kargarian2017decentralized}. Further, a numerical analysis presented in~\cite{8810817} shows that distributed optimization algorithms might converge to suboptimal solutions for nonconvex problems, depending on the initialization. These studies analyze the distributed algorithm assuming ideal communication between interconnected systems.

Communication plays a major role in implementing distributed optimization since regions share data with their neighbors~\cite{7007620}. The shared data between interconnected regions can be subject to communication errors and malicious attacks. The authors of~\cite{8240623} investigate the impact of injecting false values into the shared data of the ATC algorithm when applied to the network-constrained unit commitment problem and show that the error injection increases the convergence time. Nevertheless, it is difficult to derive a conclusion about the robustness of the algorithms when considering one type of communication nonideality with a single erroneous value injection. Reference~\cite{8113524} investigates the communication infrastructure requirements for distributed optimization when applied to power system problems. The authors of~\cite{8113524} model the communication network using a simulator and investigate the impact of communication delay on two distributed optimization algorithms. Communication delay is one important communication nonideality, but other nonidealities can impact the performance of distributed optimization algorithms.

In more general distributed optimization settings, the impacts of communication noise due to quantization effects have been investigated in~\cite{7891027,4738860,YUAN20121053}. The authors of~\cite{8629934} provide analytic upper and lower bounds on the ADMM algorithm's performance in the presence of random errors and validate the results using randomly generated networks. The authors of~\cite{9147030} study the robustness of distributed algorithms based on dual averaging to additive communication noise. The authors of~\cite{li2021distributed} investigate the ADMM algorithm's convergence under additive communication noise and recommend modifications to the underlying optimization problem. Packet loss is another communication issue that has been discussed in the literature. The impact of packet loss on unconstrained convex distributed optimization is investigated in~\cite{9264572}. A relaxed ADMM algorithm is proposed in~\cite{9198226} to optimize integrated electrical and heating systems considering communication packet loss. 

In this paper, we compare the performance of three distributed algorithms (ADMM, ATC, and APP) in solving OPF problems in the presence of nonideal communications. We characterize the algorithms' performance in terms of convergence rate and solution quality. To serve as a benchmark, we first compare the performance of the three algorithms with ideal communications. We then consider three noise models that impact the shared data between neighboring regions: (1)~additive Gaussian noise, (2)~bad data (large error), and (3)~intermittent communication loss. We use five test systems to evaluate the performance of the algorithms with nonideal communication. This paper considers the DC~OPF formulation which uses the DC power flow linearization~\cite{stott2009dcopf}. Thus, this paper's main contribution is an extensive empirical study regarding the impacts of nonideal communication on various distributed solution algorithms for DC~OPF problems.

We note that the convexity of the DC~OPF formulation studied in this paper provides advantages in terms of theoretical convergence guarantees for the distributed algorithms. These guarantees are useful in the context of this paper as they ensure that the distributed algorithms should all converge to the same solution, thus permitting consistent comparisons that primarily focus on the speed and robustness of the algorithms. However, there are applications where the DC power flow approximation is inappropriate, thus requiring alternative power flow representations~\cite{molzahn_hiskens-fnt2019}. This paper forms a basis for future extensions of our analyses for these applications.

The rest of the paper is organized as follows. Section~\ref{SEC2} introduces the mathematical notation and the DC OPF problem. Section~\ref{SEC3} formulates the problem decomposition and the distributed solution algorithms. Section~\ref{SEC4} describes the noise models we use in the analysis. Section~\ref{SEC5} presents numerical results and discusses the distributed algorithms' performance under nonideal communication. Section~\ref{SEC6} summarizes the main findings and discusses future work.

\subsection*{Remark on Notation}
Throughout the paper, we use bold letters to indicate vectors. We denote a local subproblem and regions with the letter $m$ and let $\mathcal{M}$ indicate the set of all subproblems. The set $\mathcal{N}_s^m$ contains the boundary variables for region $m$. We use the letters $n$ and $c$ to denote variables from the neighboring regions and the central coordinator. We use $||\,\cdot\,||$ to denote the vector $l_2$-norm. We use the hat in $\hat x$ to indicate that the variables $x$ are evaluated with constant values from the previous iteration as received from neighboring regions or obtained from the prior local solution.

\section{DC Optimal Power Flow Formulation} \label{SEC2}
OPF problems typically minimize generation costs while satisfying the power flow equations and limits on generator outputs, line flows, etc. The optimization variables are the generators' outputs and the bus voltages. We consider the DC power flow approximation~\cite{stott2009dcopf} to obtain convergence guarantees for the distributed algorithms, which are discussed in Section~\ref{SEC3}. The DC~OPF formulation used in this paper is:
\begin{subequations}%
\label{eq:dcopf}%
\begin{align}\label{eq:1}%
& \min_{\pmb{\theta},\pmb{p}} \quad \sum_{i\in \mathcal{G}} f_i (p_i) \\
& \text{s.t.}\quad \label{eq:2}
p_i - d_i = \sum_{(i,j)\in \mathcal{L}} B_{ij} (\theta_i - \theta_j), & \forall i\in \mathcal{B}\\
& \hphantom{\text{s.t.}}\quad \label{eq:3}
P_{i}^{min} \le p_i \le P_{i}^{max}, & \forall i \in \mathcal{G}\\
& \hphantom{\text{s.t.}}\quad \label{eq:4} -P_{ij}^{max} \le B_{ij} (\theta_i - \theta_j) \le P_{ij}^{max}, & \forall (i,j)\in \mathcal{L}\\
& \hphantom{\text{s.t.}}\quad \label{eq:5}\theta^{ref} = 0 
\end{align}
\end{subequations}
where the sets $\mathcal{B}$, $\mathcal{G}$, and $\mathcal{L}$ denote the buses, generators, and lines, respectively. The generation costs are denoted by $f_i$. The decision variables are $\pmb{\theta}$, the bus voltage angles, and $\pmb{p}$, the generators' active power outputs. We denote the power demands by $d$, and $B$ denotes the lines' susceptances. Bounds on the power output of generator~$i$ are $P_{i}^{max}$ and $P_{i}^{min}$, and $P_{ij}^{max}$ defines the flow limit of the line between buses~$i$ and~$j$. The objective function in~\eqref{eq:1} minimizes the total cost of the generators' power outputs. Constraint~\eqref{eq:2} is the DC approximation of the power flow equations~\cite{stott2009dcopf}. Constraints~\eqref{eq:3} and~\eqref{eq:4} limit the generators' power outputs and the line flows. Constraint \eqref{eq:5} sets the reference angle,~$\theta^{ref}$.

\section{Distributed Optimization Algorithms} \label{SEC3}
\begin{figure}[ht] 
    \centering
	    \includegraphics[width=3.4in]{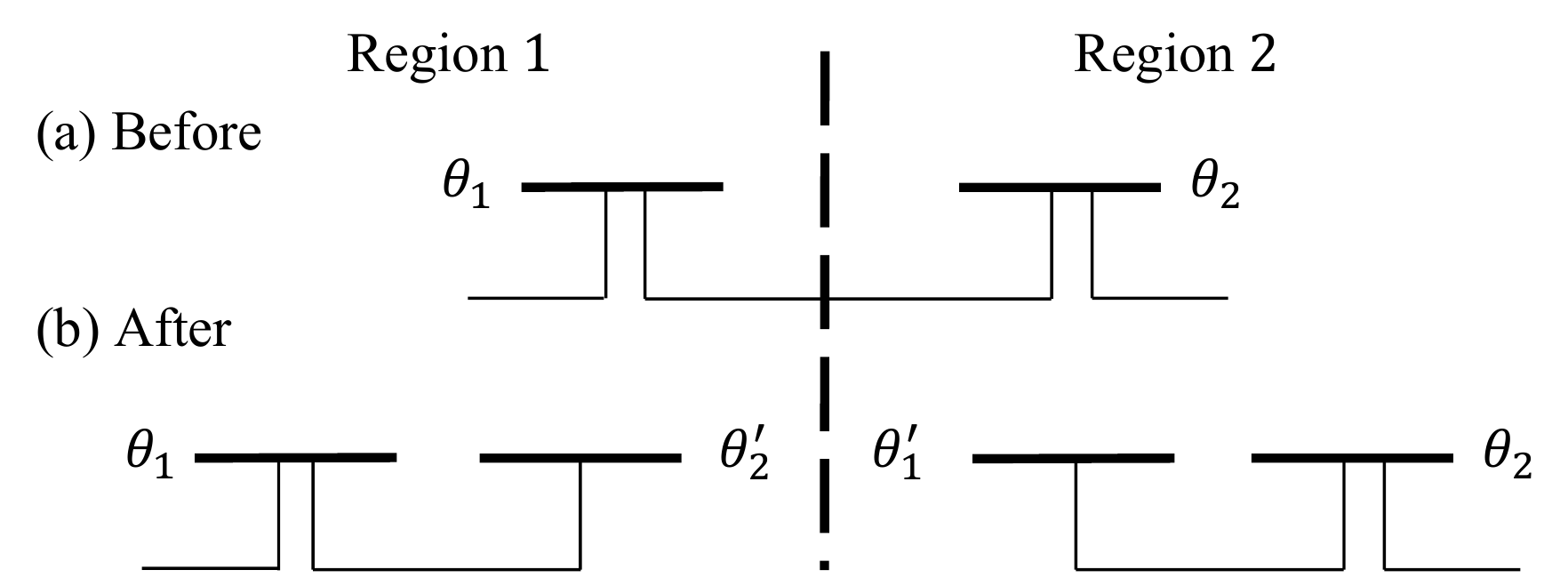}
	\caption[Decomposition]{Tie-line model before (a) and after (b) applying the decomposition.}
	\label{fig1}
\end{figure}

This section overviews the distributed algorithms considered in this paper. These algorithms decompose the centralized DC~OPF problem into subproblems corresponding to a partition of the original system. In the DC~OPF formulation~\eqref{eq:dcopf}, tie-lines between regions couple the power flow equations~\eqref{eq:2} and the line flow limits~\eqref{eq:4}. For each tie-line, we duplicate the boundary variables and assign a local copy to each region as shown in Fig.~\ref{fig1}.

To obtain a feasible solution to the centralized problem, we need to ensure consistency between the duplicated variables in each subproblem. To accomplish this, we consider algorithms that use an Augmented Lagrangian method to enforce consistency using a penalized objective function:
\begin{equation}\label{eq:al}
   L_{\rho}(\pmb{p},\pmb{\theta},\pmb{\lambda}) := \sum_{i\in \mathcal{G}} f_i (p_i) + \sum_{i\in \mathcal{N}_s}\lambda_i ( \theta_i - \theta_i^{\prime}) + \frac{\rho}{2} ||  \theta_i - \theta_i^{\prime} ||_2^2,
\end{equation}
where $\pmb{\lambda}$ are the Lagrange multipliers of the consistency constraints, $\pmb{\theta}^{\prime}$ are copies of the boundary variables, and $\rho$ is a parameter. The set $\mathcal{N}_s$ contains all copies of the bus voltage angle variables corresponding to the tie-lines between regions.

The three distributed algorithms we consider are similar in their application of an Augmented Lagrangian to decompose the original problem. However, they use different processes for evaluating the objective function and updating the Lagrange multiplier $\lambda$. Generally, the three algorithms use the shared variables received from neighboring regions to update the Lagrange multipliers and evaluate the relaxed consistency constraints in the objective function. This makes the subproblems separable and independent from each other. The algorithms repeat these steps until all regions reach a consensus on the shared variables' values. We use the $\ell_2$-norm of the shared variables' mismatch, denoted by $||\Delta\theta||$, to measure the consensus on the shared variables. We next provide more detail for these distributed algorithms.

\subsection{Alternating Direction Method of Multipliers (ADMM)}
ADMM is a well-known algorithm for solving large optimization problems~\cite{gabay1976dual,glowinski1975approximation}. The ADMM algorithm solves the augmented Lagrangian problem in a distributed fashion that is similar to the Gauss-Siedel iterative method. Solving the OPF problem using ADMM in a distributed way involves a central coordinator to decompose the problem. The local problem of region $m$ for iteration $k+1$ is:
\begin{align} \nonumber
   & \min_{\pmb{p}^{k+1},\pmb{\theta}^{m,k+1}} \sum_{i\in \mathcal{G}^m} f_i (p_i^{k+1}) + \sum_{i\in \mathcal{N}_s^m}\lambda^k_i (\hat\theta^{c,k}_i - \theta^{m,k+1}_i) \\\label{eq:admm1}
   & \qquad\qquad\qquad\qquad + \frac{\alpha}{2} ||(\hat\theta^{c,k}_i - \theta^{m,k+1}_i) ||_2^2,\\[0.5em]
   \nonumber & \text{subject to the DC~OPF constraints \eqref{eq:2}--\eqref{eq:5},}
\end{align}
where $\alpha$ is a tuning parameter. The decision variable $\pmb{\theta}^m$ includes the local variables, i.e., the voltage angles for both the local and shared buses. The variable $\pmb{\hat \theta}^c$ denotes the shared variables evaluated with values received from the central coordinator, and $\pmb{\lambda}$ denotes the Lagrange multipliers.

After solving the local problem~\eqref{eq:admm1}, the local controllers share their solutions with the central coordinator. The coordinator then solves an unconstrained optimization problem:
\begin{align}\nonumber 
    \min_{\pmb{\theta}^{c,k+1}} \: \sum_{m\in \mathcal{M}} \sum_{i\in \mathcal{N}_s^m} & \: \lambda_i (\theta^{c,k+1}_i\! - \hat \theta^{m,k+1}_i) \\
   \label{eq:admm2} 
   &+ \frac{\alpha}{2} ||(\theta^{c,k+1}_i\! - \hat\theta^{m,k+1}_i) ||_2^2,
\end{align}
where $\pmb{\theta}^c$ are the decision variables and $ \pmb{\hat \theta}^m$ are the shared variables received from the local controllers. The shared variable values obtained from the central coordinator's problem are then sent to the local controllers to update the Lagrange multipliers according to~\eqref{eq:admm3}:
\begin{equation}\label{eq:admm3}
   \pmb{\lambda}^{k+1} = \pmb{\lambda}^{k} + \alpha (\pmb{\theta}^{c,k+1} -\pmb{\theta}^{m,k+1}),
\end{equation}
where $\alpha$ is the same parameter used in the local optimization problem~\eqref{eq:admm1}. Note that $\pmb{\theta}^m$ and $\pmb{\theta}^c$ are the solutions of~\eqref{eq:admm1} and~\eqref{eq:admm2}, respectively. The local controllers then use the updated Lagrange multipliers and the central coordinator's solution to update the local problems. This process is repeated until the shared variables agree to within a specified tolerance.

There are several variants of this ADMM implementation that have been proposed for solving OPF problems. These include extensions that eliminate the need for a central coordinator~\cite{sun2013fully,erseghe2014distributed}. A proximal massage passing (PMP) method proposed in~\cite{kraning2013dynamic} also permits a fully distributed ADMM implementation. The calculations in this paper use a fully distributed implementation of the ADMM algorithm by replacing the central coordinator's problem with its first-order optimality conditions. In this implementation, each local controller uses the prior local solution and the neighboring regions' solutions to solve the central coordinator's problem locally. We also note that the ADMM algorithm has been proven to converge to the optimal solution if the subproblems are convex. See~\cite{boyd2011distributed} for more detail regarding ADMM implementations and convergence guarantees.

\subsection{Analytic Target Cascading (ATC)}
ATC is another distributed algorithm based on augmented Lagrangian relaxation that solves a large optimization problem via dividing it into hierarchically connected subproblems with multiple levels~\cite{michelena2003convergence}. Two levels of subproblems are connected if they share coupling variables. To solve the OPF problem, we use a two-level ATC structure. The first level consists of a central coordinator, while the regions' local problem are in the second level. The local problem $m$ for iteration $k+1$ is:
\begin{align}\nonumber
   & \min_{\pmb{p}^{k+1},\pmb{\theta}^{m,k+1}} \sum_{i\in \mathcal{G}^m} f_i (p_i) + \sum_{i\in \mathcal{N}_s^m}\lambda_i (\hat\theta^{c,k}_i - \theta^{m,k+1}_i) \\ \label{eq:ATC1}
   & \qquad\qquad\qquad\qquad + || \beta (\hat\theta^{c,k}_i - \theta^{m,k+1}_i) ||_2^2,\\[0.5em]
   \nonumber & \text{subject to DC OPF constraints \eqref{eq:2}--\eqref{eq:5},}
\end{align}
where $\pmb{\lambda}$ are the Lagrange multipliers and $\beta$ is a parameter. We denote the shared variables that are fixed to their values from the central coordinator with $\pmb{\hat\theta}^{c}$. The local controllers communicate the resulting shared variable values with a central coordinator. The coordinator solves an unconstrained optimization problem that minimizes the differences between the boundary variables for neighboring regions:
\begin{align}\nonumber
   \min_{\pmb{\theta}^{c,k+1}} \sum_{m\in \mathcal{M}} \sum_{i\in \mathcal{N}_s^m}&\lambda_i (\theta^{c,k+1}_i - \hat \theta^{m,k+1}_i) \\
    \label{eq:ATC2}
   &+ || \beta (\theta^{c,k+1}_i - \hat \theta^{m,k+1}_i) ||_2^2.
\end{align}

The coordinator shares the target results $\pmb{\theta}^{c,k+1}$ with the local controllers. Next, the local controllers update the Lagrange multipliers and the parameters $\beta$ using the target variables:
\begin{subequations}
\begin{align}\label{eq:ATC3}
   &\pmb{\lambda}^{k+1} = \pmb{\lambda}^{k} + 2 (\beta^{k})^2 (\pmb{\theta}^{c,k+1} - \pmb{\theta}^{m,k+1}),\\
   \label{eq:ATC4}
   &\beta^{k+1} = \alpha \beta^{k},
\end{align}
\end{subequations}
where $\beta$ is the same parameter used in the local optimization problem~\eqref{eq:ATC1}, and $\alpha$ is a tuning parameter. After updating the Lagrange multipliers, the local controllers and the central coordinator repeat the process until the shared variables are within a specified tolerance.

Variants of ATC use different functions besides the Augmented Lagrangian to relax the consistency constraints~\cite{tosserams2006augmented}. Further, the ATC variant proposed in~\cite{8240623} is fully distributed, eliminating the need for a central coordinator. Similar to the ADMM algorithm, we use a fully distributed implementation, where each local controller uses the first-order optimality conditions of the central coordinator's problem to solve the local subproblem. The ATC algorithm is proven to converge for convex problems if the interaction is limited to subproblems in different levels~\cite{michelena2003convergence}.

\subsection{Auxiliary Problem Principle (APP)}
The APP algorithm is also based on augmented Lagrangian decomposition~\cite{cohen1980auxiliary}. The APP algorithm solves a sequence of auxiliary problems in a distributed fashion without the need for a central coordinator. In contrast to ADMM and ATC which directly use the Augmented Lagrangian, APP linearizes the quadratic term in the augmented Lagrangian around the previous iteration and introduces a regularization term in the objective function~\cite{zhao2015auxiliary}. Using the APP algorithm, the OPF formulation for region $m$ and iteration $k+1$ is:
\begin{align} \nonumber
   &\min_{\pmb{p}^{m,k+1},\pmb{\theta}^{k+1}} \quad \sum_{i\in \mathcal{G}} f_i (p^{k+1}_i) +
   \sum_{i\in \mathcal{N}_s^m} \frac{\beta}{2} ||\theta^{m,k+1}_i - \hat \theta^{n,k}_i ||^2_2 \\ \label{eq:APP1}
   & \qquad\qquad\qquad + \gamma \theta^{m,k+1}_i (\hat \theta^{m,k}_i - \hat \theta^{n,k}_i) + \lambda \theta^{m,k+1}_i,\\
\nonumber & \text{subject to DC OPF constraints \eqref{eq:2}--\eqref{eq:5},}
\end{align}
where $\alpha$, $\beta$, and $\gamma$ are tuning parameters. We define $\pmb{\hat \theta}^{m,k}$ as the value of the shared variable obtained by region~$m$ in the previous iteration, and $\pmb{\hat \theta}^{n,k}$ denotes the values of the shared variables received from neighboring regions. After each region's local controller solves its associated problem and exchanges the results with the neighboring regions, each local controller updates the values of the Lagrange multipliers as follows:
\begin{equation}\label{eq:APP2}
   \pmb{\lambda}^{k+1} = \pmb{\lambda}^{k} + \alpha (\pmb{\theta}^{m,k+1} - \pmb{\theta}^{n,k+1}).
\end{equation}

To summarize, each region iteratively solves~\eqref{eq:APP1}, shares the results with neighboring regions, and updates the Lagrange multiplier using~\eqref{eq:APP2} until reaching consensus on the shared variables. If the local problems are convex and differentiable, selecting parameters satisfying the condition $\alpha < 2\gamma < \beta$ guarantees that the APP algorithm will converge~\cite{kim1997coarse}.

\section{Nonideal Communications Models} \label{SEC4}
The communication requirements for a distributed algorithm depend on the shared variables. During each iteration, each region shares the results of its local optimization by communicating with the neighboring regions. Since communication networks are not ideal, the shared data may suffer from data quality issues or interruptions that impact the distributed algorithms' performance. In this section, we introduce three models for nonideal communications: additive Gaussian noise, bad data (large errors), and intermittent loss of communication.

\subsection{Additive Gaussian Noise} \label{nt1}
The data shared between connected regions may be subject to noise resulting from imperfect communication. This noise could be due to quantization error~\cite{1424312} or added to enforce data privacy requirements~\cite{7563366}. To model noisy shared data, we inject additive Gaussian noise into the shared variables: 
\begin{equation} \label{nt:1}
\pmb{\theta}_{noisy} = \pmb{\theta}_{noiseless} + \pmb{N}(0, \sigma_{noise})
\end{equation}
where $\pmb{\theta}_{noisy}$ is the data that is actually communicated to the neighbors, $\pmb{\theta}_{noiseless}$ is the true data, and $\pmb{N}(0,\sigma_{noise})$ is a vector of normally distributed random numbers with zero mean and standard deviation of $\sigma_{noise}$.

\subsection{Bad Data} \label{nt2}
Neighboring regions may occasionally receive ``bad data'', i.e., data with large errors. Bad data may be due to an instantaneous bit error~\cite{1146031} or a malicious adversarial agent~\cite{6102368}. We model a random injection of bad data at a specified occurrence probability as shown in the following model:
\begin{align} \nonumber 
&\qquad \theta_{noisy} = \theta_{noiseless} + \; 2 \; R \; r, \\
\label{nt:2}
&\text{where} \quad r = \begin{cases} 
U_1-0.5 &\text{if}\;\; U_2 < p, \\
0   &\text{otherwise}.
\end{cases}
\end{align}

In~\eqref{nt:2}, $R$ is the maximum magnitude of the error and $r$ is the error multiplier that randomly selects the error magnitude. The variables $U_1$ and $U_2$ are uniformly distributed random numbers between $[0, 1]$, and $p$ is the probability of bad data occurrence per iteration. 

\subsection{Intermittent Communication Loss} \label{nt3}
Communications between the agents may occasionally fail entirely for multiple iterations. For instance, communication transmission collisions or instability of the communication link can cause packet loss preventing the agents from sharing data for a number of iterations~\cite{9264572}. We consider a simple two-state model with success and fail states to represent the loss of communication. If the communication channel is in a success state, then the data will be transmitted while the fail state means the data will be lost. The transition from one state to another occurs with a constant probability. This model is similar to Gilbert-Elliott Erasure channel used to model pocket loss~\cite{hasslinger2008gilbert}. Although communication loss can have more complicated occurrence behaviour, this model is used to model pocket loss due to its tractability and reasonable behaviour with respect to experimental data~\cite{1223556}. 

To model intermittent communication loss, we define the \emph{failure probability}, denoted $\lambda_f$, as the transition probability of the communication channel from success to fail states per iteration given that the channel was in success state during the previous iteration. Similar to the failure probability, we define the \emph{repair probability}, denoted $\lambda_r$, as the transition probability from fail to success state. We also introduce a state variable $s$ for each communication channel connecting two regions, where $s = 1$ is a success state and $s = 0$ is a fail state. We use an indicator function $\pmb{1}\{x\}$, which equals 1 if $x$ is true and 0 otherwise. We model the state of the communication channel by sampling from a uniformly distributed random number $U$ at each iteration and compare it with the failure (repair) probability to decide the channel state. If the controller detects an interruption from a neighboring region, it uses the value from the last successful data transmitted from this region. The following pseudocode describes our intermittent communication loss model:
\begin{algorithm} 
    \renewcommand{\thealgorithm}{}
    \floatname{algorithm}{}
    \caption{Intermittent Communication Loss Model} 
    \begin{algorithmic}[1]
        \STATE {$\text{generate a random number}$ $U \in [0,1]$}
        \STATE \algorithmicif {\;\; $s = 1$\;} \algorithmicthen {$\;\; \text{set} \;\; s = 1 - \pmb{1} \{U < \lambda_f\}$}
        \STATE \algorithmicelse {$\;\; \text{set} \;\;  s = \pmb{1} \{U < \lambda_r\}$} \algorithmicendif
        \STATE \algorithmicif {\;\; $s =0$\;} \algorithmicthen$\;\; \text{set} \;\; \pmb{\theta}_{shared}^{k+1} = \pmb{\theta}_{shared}^{k}$  \algorithmicendif
    \end{algorithmic}
\end{algorithm}

\section{Performance Analyses} \label{SEC5}
This section presents the numerical results of solving the DC~OPF problem using the three distributed algorithms. We first present the results of each algorithm with ideal communication. We then compare the performance of the distributed algorithms with the nonideal communication models in Section~\ref{SEC4}. We use five test systems: the 5-bus system ``WB5'' from~\cite{6581918} and the IEEE 14-bus, RTS~GMLC 73-bus, IEEE 118-bus, and IEEE 300-bus systems from M{\sc atpower}~\cite{5491276}. The first two systems are decomposed into two regions, while the latter three systems are decomposed into three regions. We model the distributed algorithm using a combination of JuMP~\cite{DunningHuchetteLubin2017} and PowerModels~\cite{8442948} libraries in Julia programming language~\cite{bezanson2017julia}, and use Gurobi to solve the subproblems.

\subsection{Performance with Ideal Communications}
We use the distributed algorithms assuming each region shares the exact results with neighboring regions at each iteration. We use the two-norm of the mismatches between the values of the shared variables to measure the consensus and set the stopping criteria. We use the results of the ideal communication to tune the parameters and evaluate the convergence rates of the algorithms. 

\subsubsection{Alternating Direction Method of Multipliers (ADMM)}

\begin{figure}[h]
    \centering
	    \includegraphics[width=3.4in]{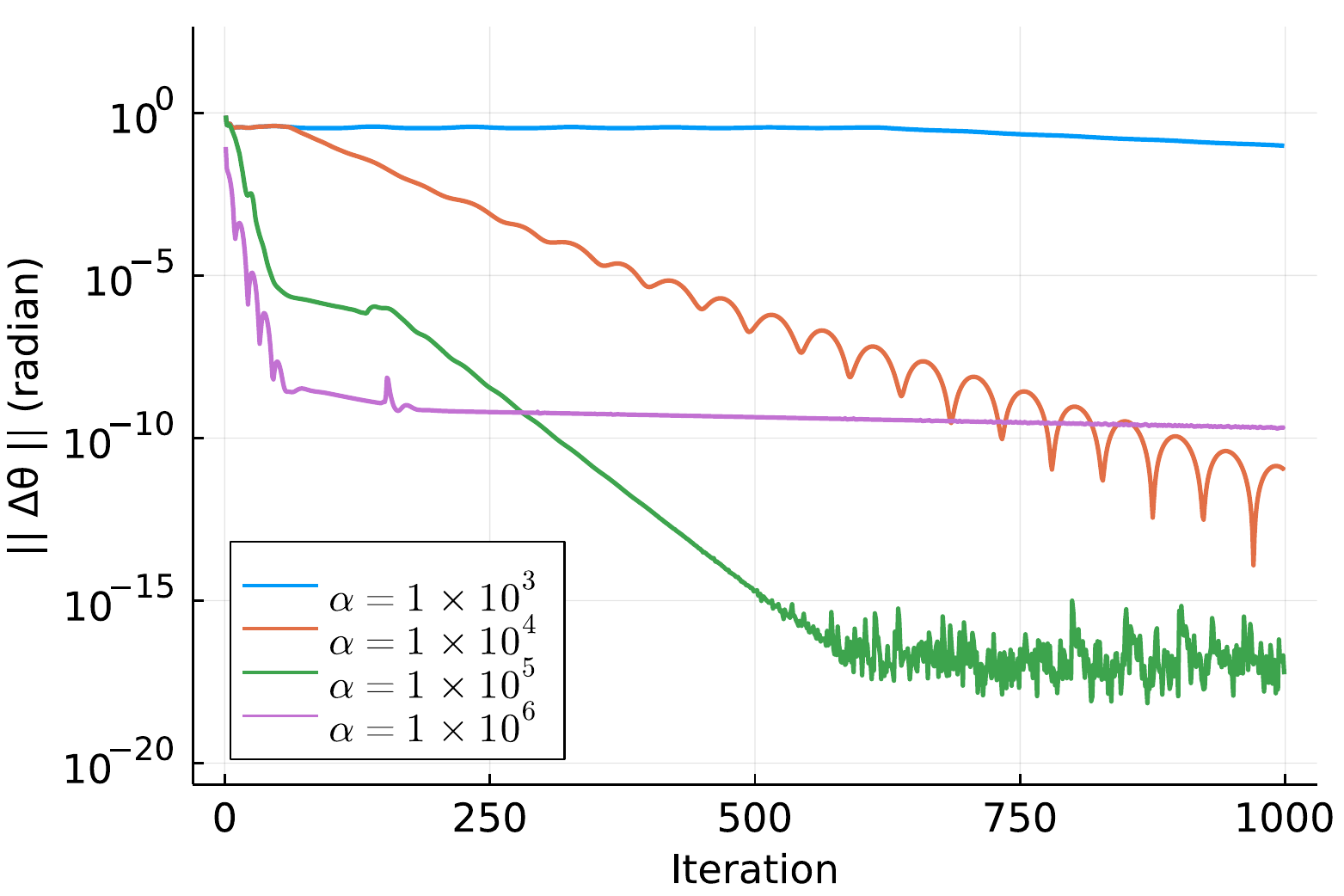}
	\caption[ADMM-Ideal]{ADMM convergence with different parameters for the IEEE 118-bus system with ideal communications.}
	\label{fig:ADMM_parameter}
\end{figure}

The ADMM algorithm's performance depends on the value selected for the parameter $\alpha$. Fig.~\ref{fig:ADMM_parameter} shows the convergence of the ADMM algorithm for the IEEE 118-bus system. For this test system, we observe that the regions reach consensus on the shared variables the fastest when the value of $\alpha$ is $1\times10^6$ at the first 50 iterations. However, we obtain more stable convergence when we set $\alpha$ equal $1\times10^5$ and the convergence became faster after 260 iterations. For the WB5, IEEE 14-bus, RTS~GMLC, and IEEE 300-bus systems, we tune the value of $\alpha$ to be $1\times10^2$, $1\times10^3$, $1\times10^5$ and $1\times10^5$, respectively. Selecting smaller values for $\alpha$ increases the number of iterations to achieve consensus, while selecting significantly larger values cause oscillations that reduces the convergence rate.

\subsubsection{Analytical Target Cascading (ATC)}
\begin{figure}[h]
    \centering
	    \includegraphics[width=3.4in]{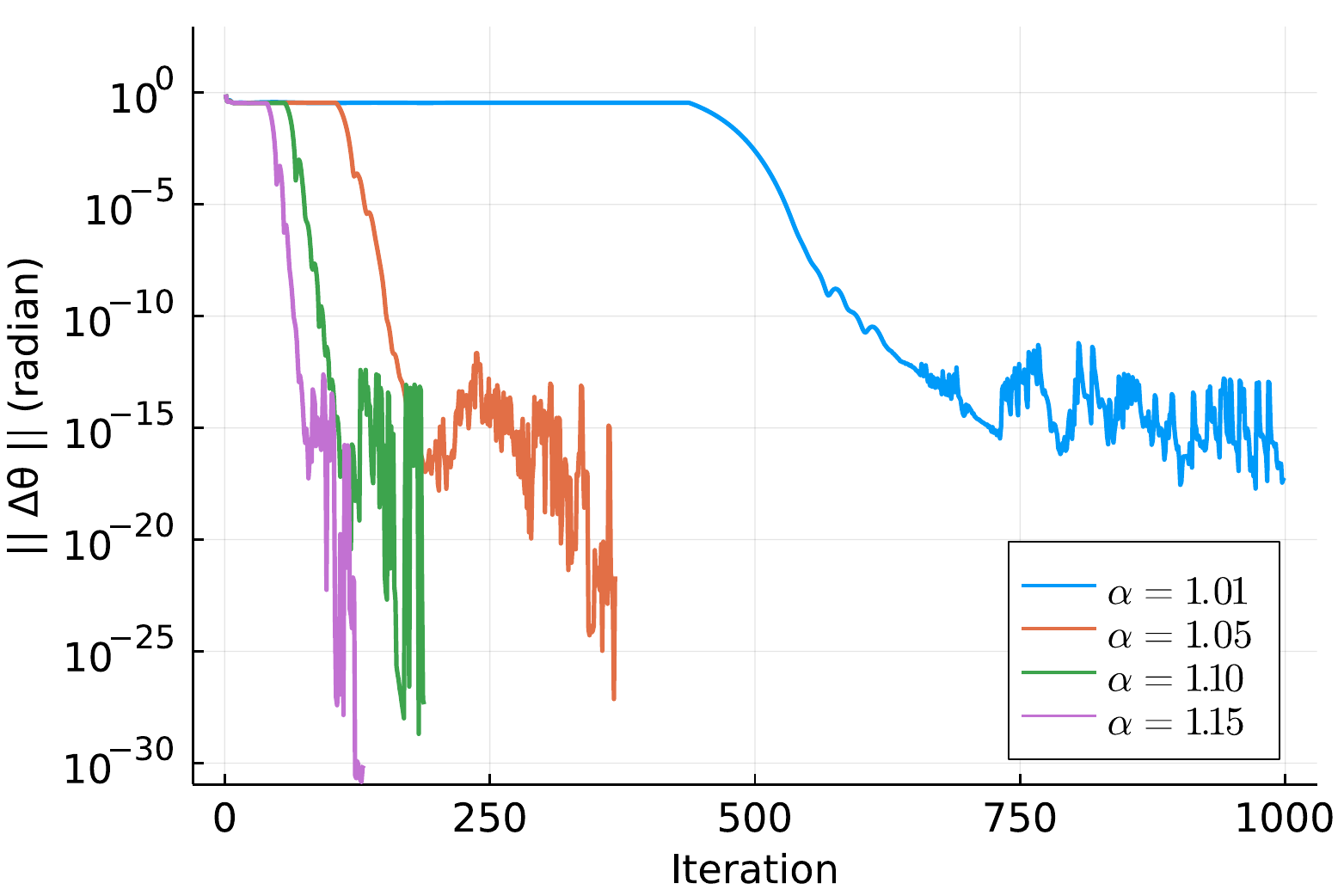}
	\caption[ATC-Ideal]{ATC convergence with different parameters for the IEEE 118-bus system with ideal communications.}
	\label{fig:ATC_parameter}
\end{figure}

Similar to ADMM, the ATC algorithm requires tuning one parameter $(\alpha)$. The convergence of the shared variables for the IEEE 118-bus system is shown in Fig.~\ref{fig:ATC_parameter}. We observe that setting the parameter $\alpha = 1.1$ increases the convergence rate by a factor of two compared to $\alpha = 1.05$. We see similar behaviour for the other systems. Furthermore, the solver fails to find an optimal solution to the local problems when the number of iterations increases as shown in~Fig.~\ref{fig:ATC_parameter}. This behaviour occurs due to the parameter update criteria~\eqref{eq:ATC4}, which exponentially increases the penalty on the shared variable consistency term as the number of iterations increases.

\subsubsection{Auxiliary Problem Principal (APP)}
\begin{figure}[h]
    \centering
	    \includegraphics[width=3.4in]{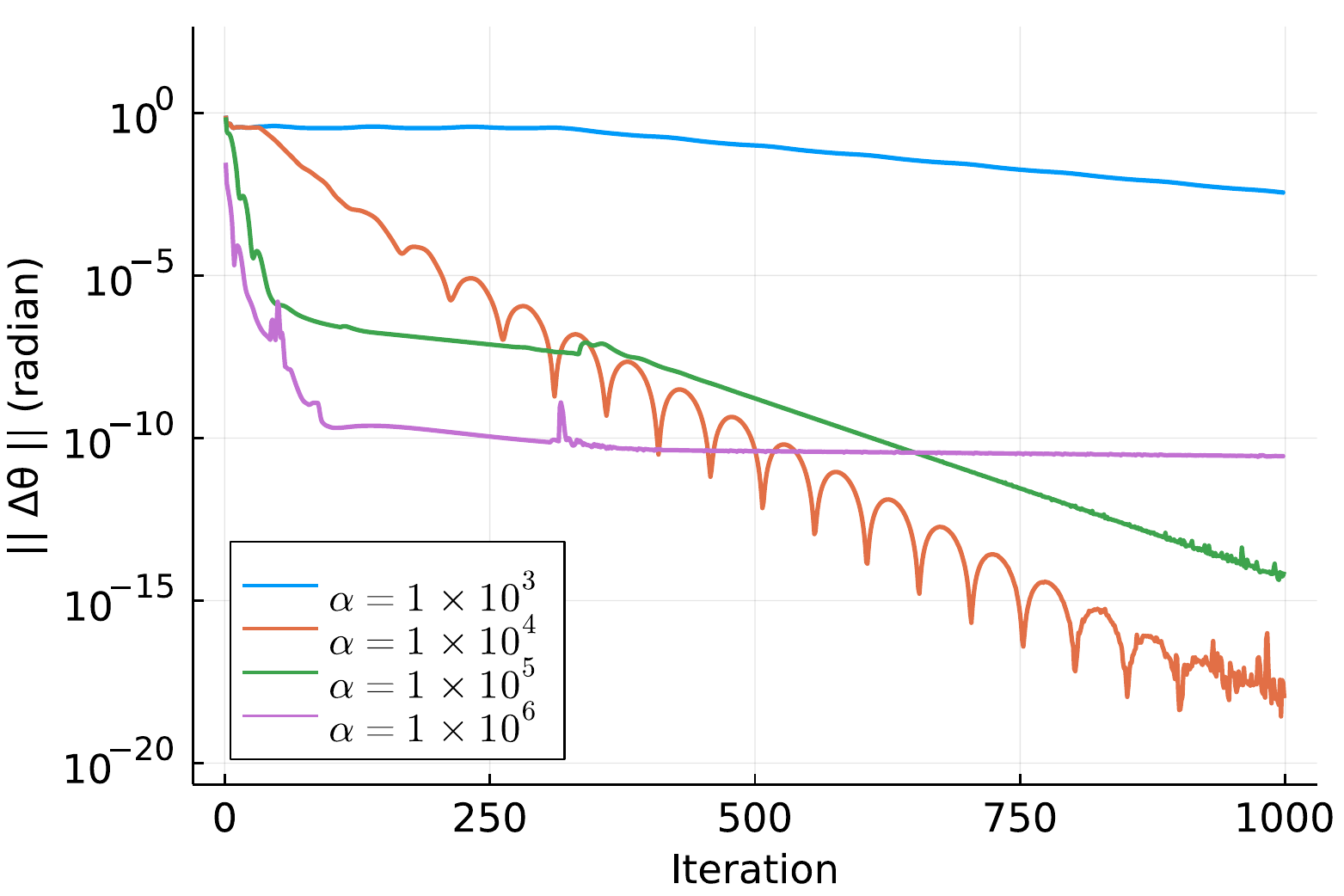}
	\caption[APP-Ideal]{APP convergence with different parameters for the IEEE 118-bus system with ideal communications.}
	\label{fig:APP_parameter}
\end{figure}

Unlike ADMM and ATC, the APP algorithm contains three parameters that need to be tuned. We adopt the condition $\alpha = \gamma = \frac{1}{2} \beta$ from prior literature~\cite{kim1997coarse} in order to simplify the parameter tuning. For the IEEE~118-bus system with three regions, the APP algorithm converges when $\alpha=1\times10^4$ as shown in Fig.~\ref{fig:APP_parameter}. The algorithm does not converge with the value of $\alpha=1\times10^3$ after 1000 iterations. For the WB5, IEEE~14-bus, RTS~GMLC, and IEEE~300-bus systems, we find a value of $\alpha$ equal to $1\times10^2$, $1\times10^3$, $1\times10^5$, and $1\times10^5$, respectively, yield a fast convergence rate. While selecting larger parameter values might improve the convergence rate, this also degrades the mismatch error in the shared variables achieved by the algorithm.

\subsubsection{Remark about the Algorithms' Performance}
Along with the convergence rate, the quality of the final solution is another metric for selecting parameters. To evaluate solution quality, we use the~\emph{Relative Gap} (RG), defined as the absolute value of the difference between the generation cost of the distributed and centralized solutions divided by the centralized solution's cost. Table~\ref{Tb1} summarizes the convergence rate results obtained from the three algorithms after reaching consensus on the shared variables with tolerance equal to $1\times 10^{-4}$ radians ($5.7\times 10^{-3}$ degrees). The convergence rate of the distributed algorithms depends on the selection of the tuning parameters. Comparing the convergence rate of the three algorithms with the tuning parameters that we select, no algorithm outperforms the others, and the fastest algorithm varies depending on the test system.
\begin{table}[h]
\caption{Convergence Results with Ideal Communication}
\label{Tb1}
\centering
\begin{tabular}{|m{1.5cm}|m{1.5cm}||c|c|c|}
\hline
\textbf{System} & \textbf{Measure} & \multicolumn{1}{c|}{\textbf{ADMM}} & \multicolumn{1}{c|}{\textbf{ATC}} & \multicolumn{1}{c|}{\textbf{APP}} \\ \hline\hline
\multirow{4}{*}{\textbf{WB5}} & $\mathbf{\alpha}$ & $10^3$ & $1.5$ & $10^2$ \\ \cline{2-5} 
 & \textbf{Iterations} & $61$ & $22$ & $34$ \\ \cline{2-5}
 & \textbf{Time (s)} & $0.1631$ & $0.0642$ & $0.0894$ \\ \hline\hline
\multirow{4}{*}{\textbf{14-Bus}} & $\mathbf{\alpha}$ & $10^4$ & $1.04$ & $10^5$ \\ \cline{2-5} 
 & \textbf{Iterations} & $116$ & $149$ & $107$ \\ \cline{2-5}
 & \textbf{Time (s)} & $0.3134$ & $0.4064$ & $0.2897$ \\ \hline\hline
\multirow{4}{*}{\textbf{RTS}} & $\mathbf{\alpha}$ & $10^7$ & $1.3$ & $10^7$ \\ \cline{2-5}
 & \textbf{Iterations} & $33$ & $35$ & $33$ \\ \cline{2-5}
 & \textbf{Time (s)} & $0.8871$ & $1.0216$ & $0.9153$ \\ \hline\hline
\multirow{4}{*}{\textbf{118-Bus}} & $\mathbf{\alpha}$ & $10^6$ & $1.1$ & $10^6$ \\ \cline{2-5} 
 & \textbf{Iterations} & $55$ & $87$ & $62$ \\ \cline{2-5}
 & \textbf{Time (s)} & $0.4973$ & $0.8381$ & $0.5842$ \\ \hline\hline
\multirow{4}{*}{\textbf{300-Bus}} & $\mathbf{\alpha}$ & $10^7$ & $1.2$ & $10^7$ \\ \cline{2-5}
 & \textbf{Iterations} & $130$ & $62$ & $100$ \\ \cline{2-5}
 & \textbf{Time (s)} & $2.1788$ & $1.0833$ & $1.7360$ \\ \hline
\end{tabular}
\end{table}
\subsection{Performance with Additive Gaussian noise}

\begin{figure}[h]
    \centering
        \includegraphics[width=3.3in]{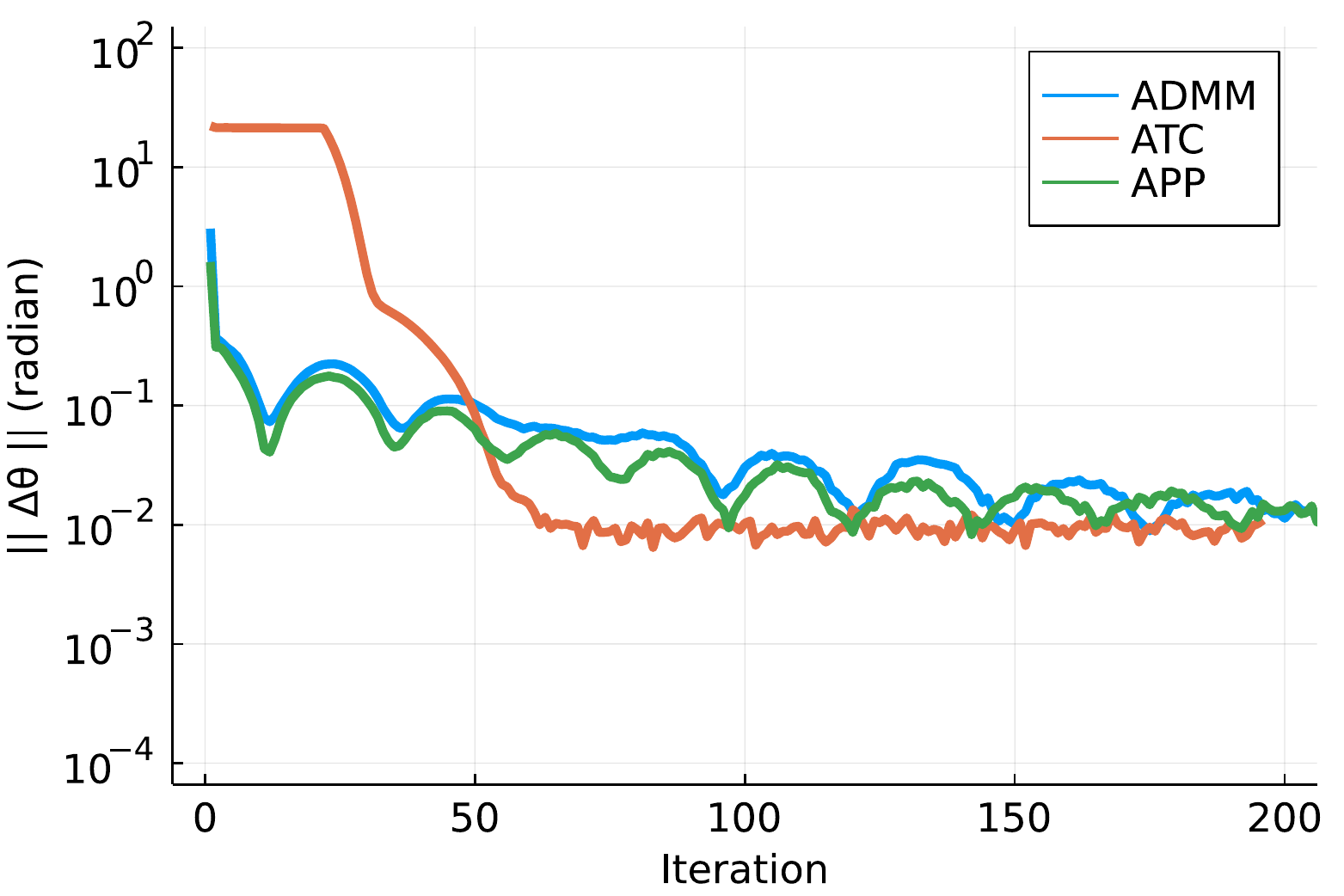}
	\caption[nt1convergence]{The three algorithms' convergence for the IEEE 118-bus system with additive Gaussian noise ($\sigma_{noise} = 1\times 10^{{-}3}$).}
	\label{fig:nt1_1}
	\vspace{-5pt}
\end{figure}

We next assume the regions send shared variable information with additive Gaussian noise as described in~\eqref{nt:1}. To compare the performance of the algorithms, we vary the standard deviation of the noise $\sigma_{noise}$. The consensus on the voltage angles achieved by the three algorithms for the IEEE 118-bus system with standard deviation $\sigma = 1\times10^{-3}$ radians is shown in Fig.~\ref{fig:nt1_1}. The results indicate that small noise levels do not significantly impact the convergence rate of the algorithms. Further, the three distributed algorithms converge to a similar level of accuracy as the level of the injected noise. However, the ATC algorithm can exhibit numerical instability preventing the solver from solving the subproblems when the algorithm is not terminated, which we attribute to the parameter update rule~\eqref{eq:ATC4}.

\begin{figure}[h]
    \centering
        \includegraphics[width=3.3in]{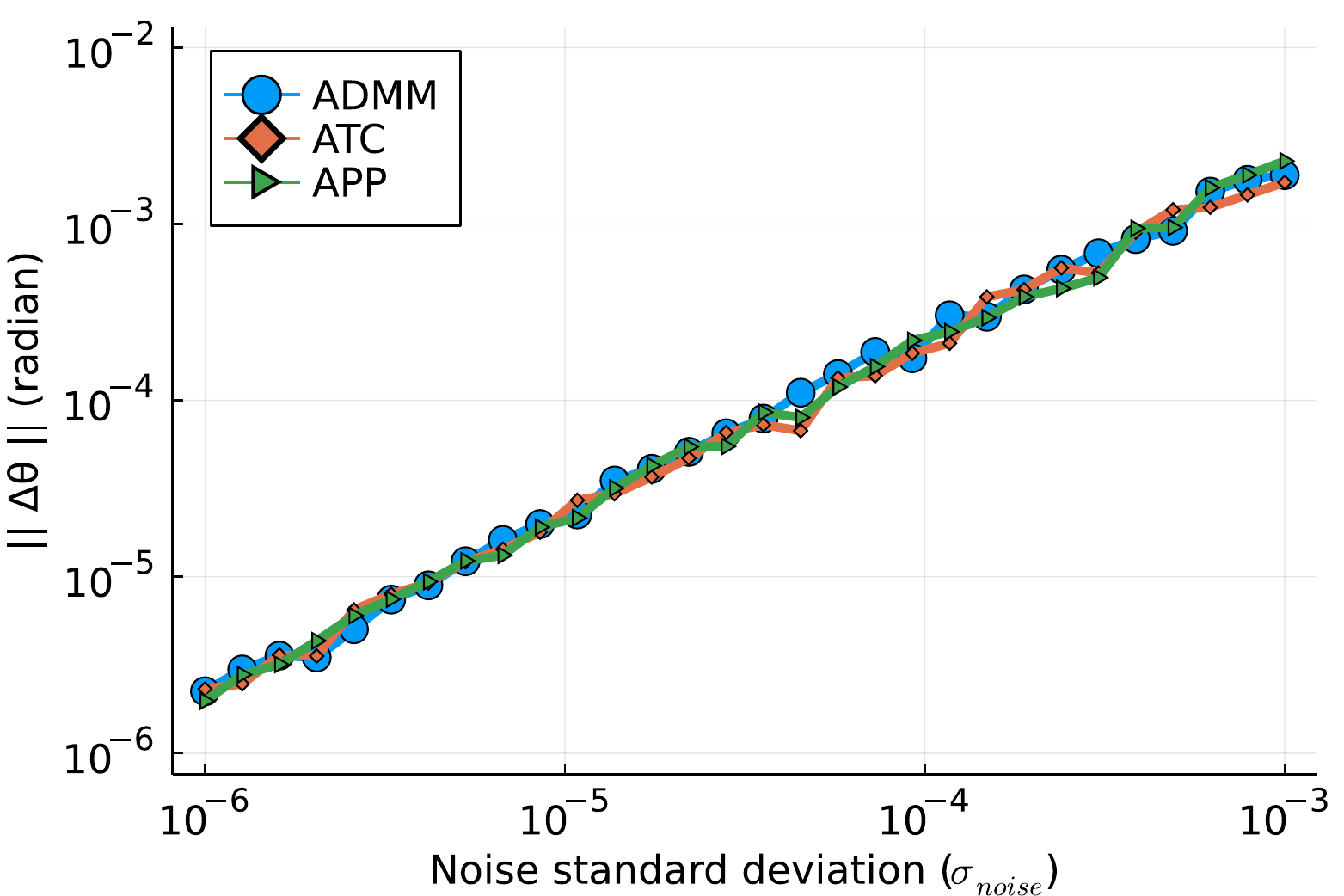}
	\caption[nt1comparison]{Mean of the mismatch in the final solution for the WB5 system with different noise levels.}
	\label{fig:nt1_mean_5}
	\vspace{-1pt}
\end{figure}

\begin{figure}[h]
    \vspace{-1pt}
    \centering
        \includegraphics[width=3.3in]{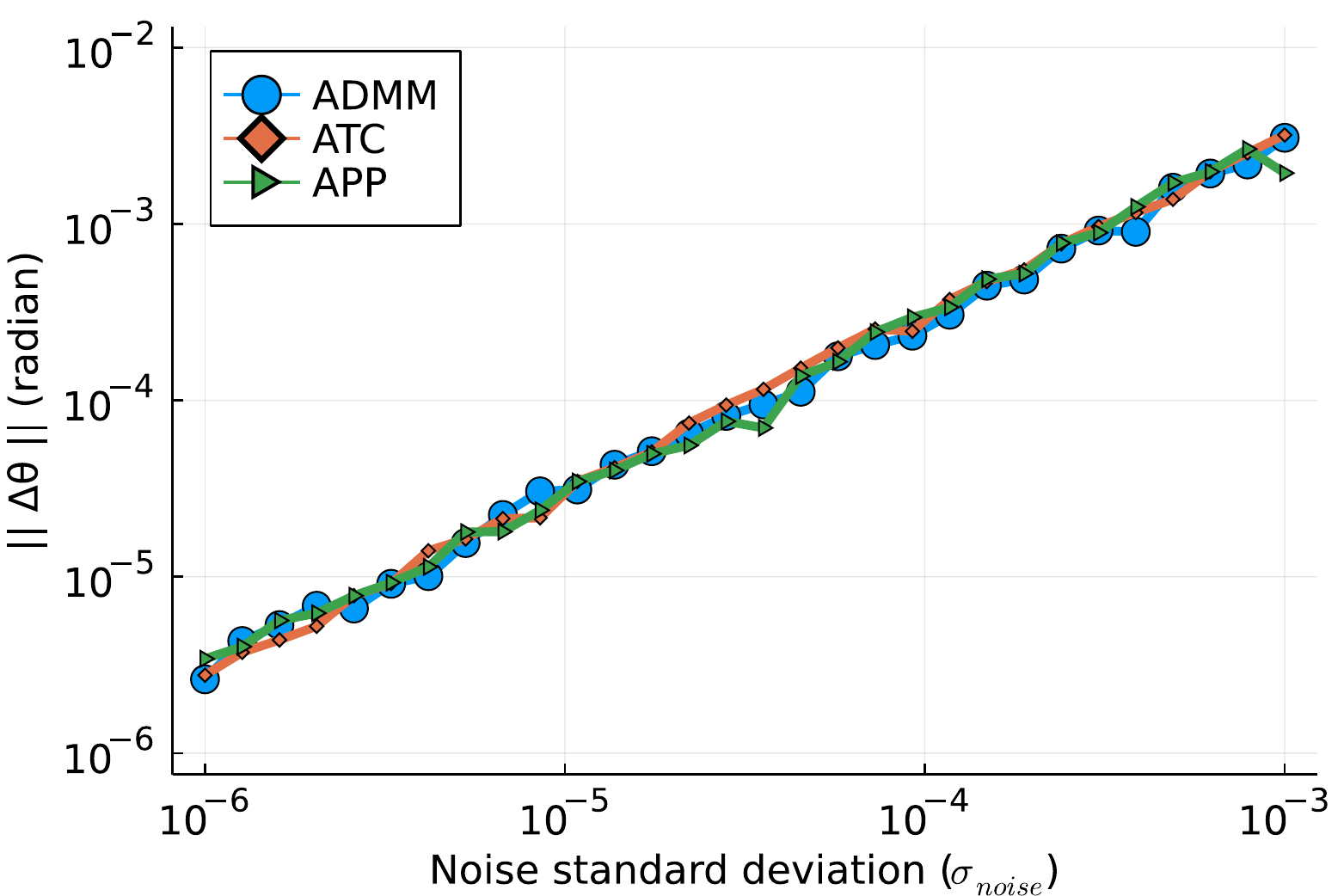}
	\caption[nt1comparison]{Mean of the mismatch in the final solution for the IEEE 14-bus system with different noise levels.}
	\label{fig:nt1_mean_14}
\end{figure}

\begin{figure}[h]
    \centering
        \includegraphics[width=3.3in]{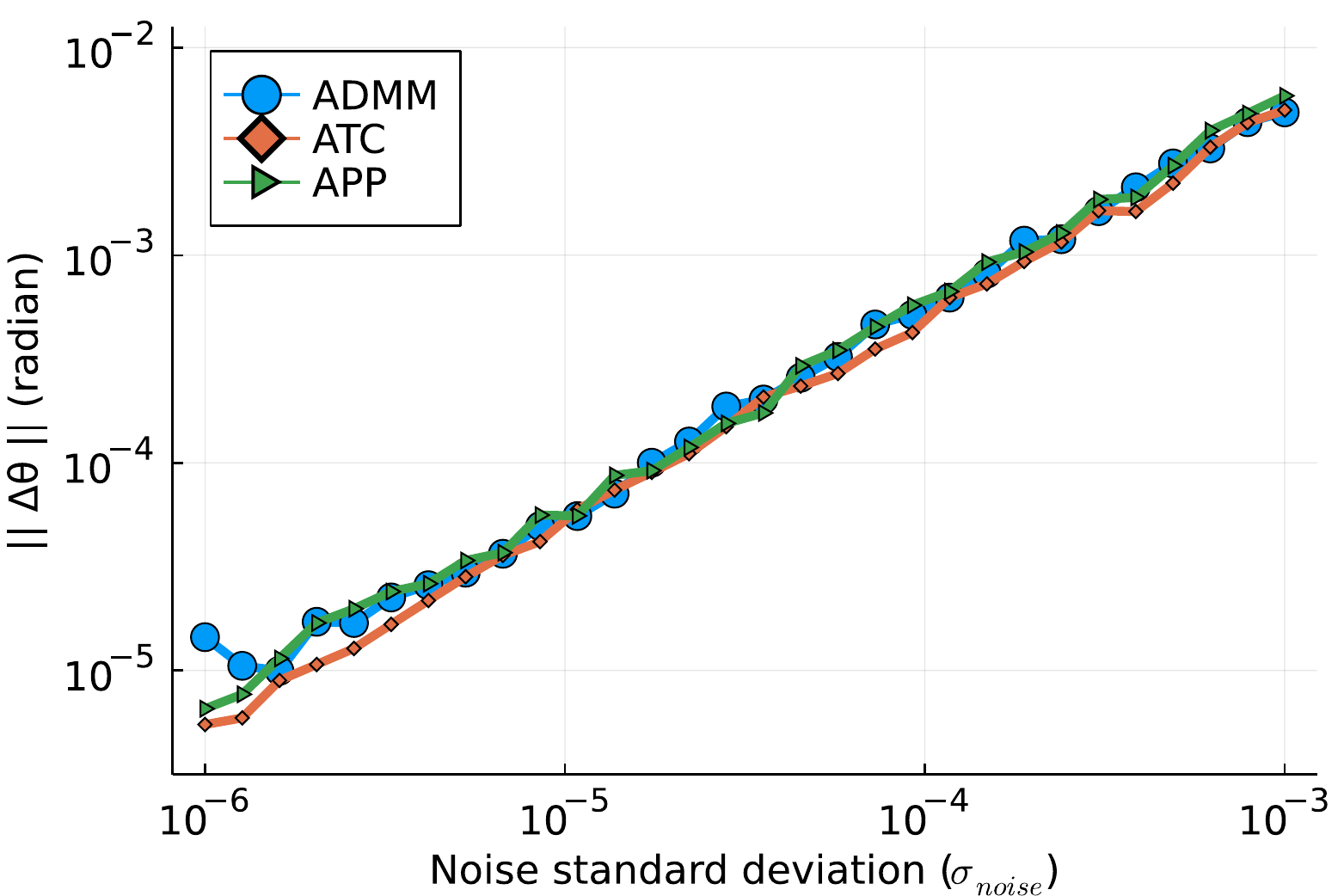}
	\caption[nt1comparison]{Mean of the mismatch in the final solution for the RTS GMLC system with different noise levels.}
	\label{fig:nt1_mean_rts}
\end{figure}

\begin{figure}[h]
    \centering
        \includegraphics[width=3.3in]{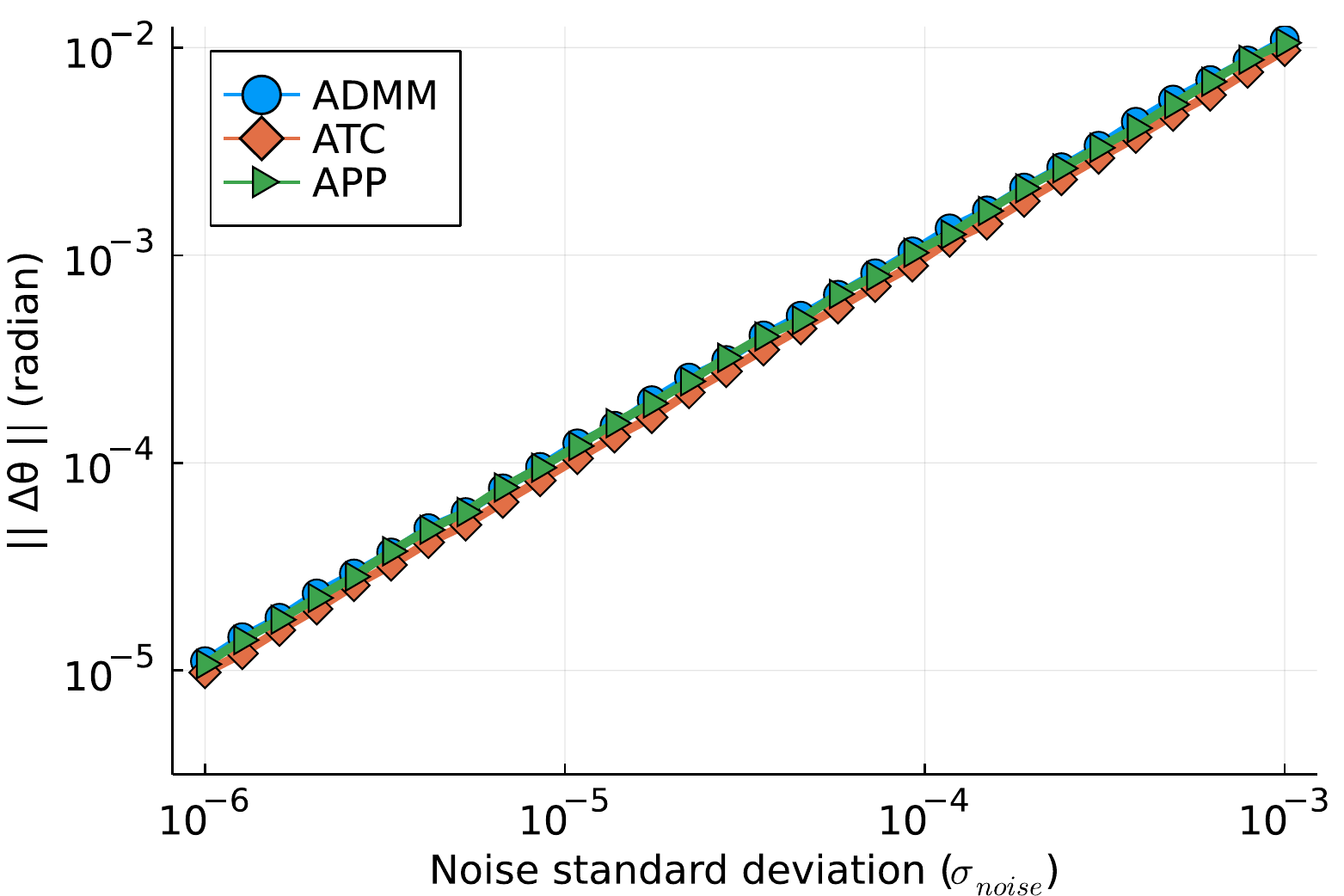}
	\caption[nt1comparison]{Mean of the mismatch in the final solution for the IEEE 118-bus system with different noise levels.}
	\label{fig:nt1_mean_118}
\end{figure}

\begin{figure}[h]
    \centering
        \includegraphics[width=3.3in]{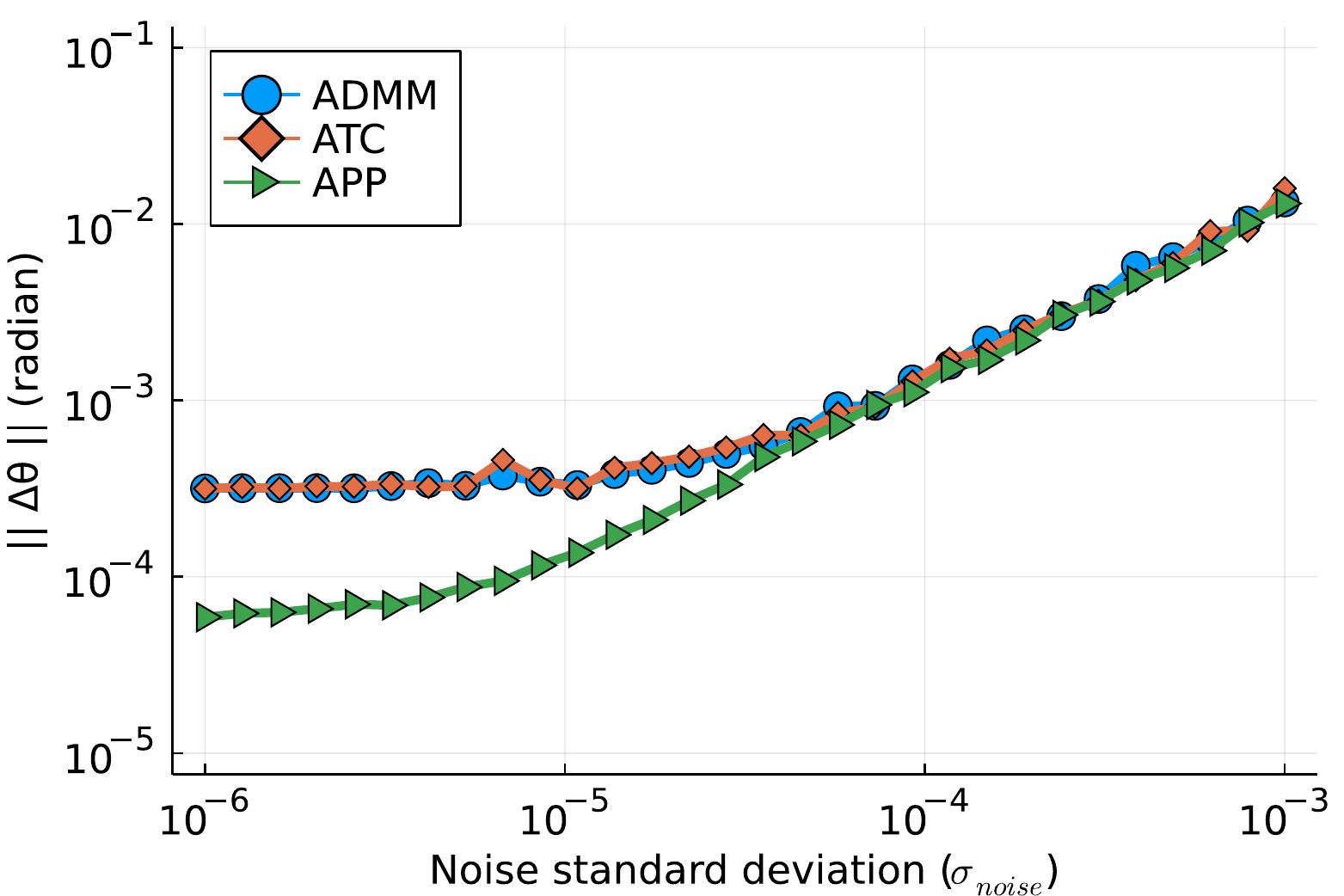}
	\caption[nt1comparison]{Mean of the mismatch in the final solution for the IEEE 300-bus system with different noise levels.}
	\label{fig:nt1_mean_300}
\end{figure}

We run the distributed algorithms 100 times and calculate the mean and the standard deviation of the shared variables mismatch. The mean and the standard deviation of the shared variables' mismatches in the final solution, $\mu_{|| \Delta\theta ||}$ and $\sigma_{|| \Delta\theta ||}$, for the three algorithms with three noise levels are shown in Table~\ref{Tb2}. To visualize the performance differences, Fig.~\ref{fig:nt1_mean_5}-~\ref{fig:nt1_mean_300} show the mean of the shared variable mismatches, $\mu_{|| \Delta\theta ||}$, in the final solutions for the test systems as the noise standard deviation, $\sigma_{noise}$, varies from $1\times10^{-6}$ to $1\times10^{-3}$. The three algorithms achieve consensus on the shared variables with an error that increases approximately linearly with the standard deviation of the added noise, with a slightly lower mismatch observed for ATC compared to ADMM and APP. This is specially noticeable for the IEEE 300-bus case with low noise standard deviating.
\begin{table*}[t!]
\centering
\caption{Mean ($\mathbf{\mu}_{|| \Delta\theta ||}$) and standard deviation ($\mathbf{\sigma}_{|| \Delta\theta ||}$) of the mismatch in the final solution with different values of noise ($\sigma_{noise}$)}
\label{Tb2}
{\scriptsize
\begin{tabular}{|l|l||r|r|r||r|r|r||r|r|r|}
\hline
\multicolumn{2}{|c||}{\textbf{Algorithm}} & \multicolumn{3}{c||}{\textbf{ADMM}} & \multicolumn{3}{c||}{\textbf{ATC}} & \multicolumn{3}{c|}{\textbf{APP}} \\ \hline
\multicolumn{1}{|c|}{\textbf{System}} & \multicolumn{1}{c||}{$\mathbf{\sigma_{noise}}$} & \multicolumn{1}{c|}{$\mathbf{10^{-5}}$} & \multicolumn{1}{c|}{$\mathbf{10^{-4}}$} & \multicolumn{1}{c||}{$\mathbf{10^{-3}}$} & \multicolumn{1}{c|}{$\mathbf{10^{-5}}$} & \multicolumn{1}{c|}{$\mathbf{10^{-4}}$} & \multicolumn{1}{c||}{$\mathbf{10^{-3}}$} & \multicolumn{1}{c|}{$\mathbf{10^{-5}}$} & \multicolumn{1}{c|}{$\mathbf{10^{-4}}$} & \multicolumn{1}{c|}{$\mathbf{10^{-3}}$} \\ \hline\hline
\multirow{2}{*}{\textbf{WB5}} & $\mathbf{\mu}_{||\Delta \theta ||}$ & $4.8\times 10^{-5}$ & $4.3\times 10^{-4}$ & $4.6 \times 10^{-3}$ & $4.4\times 10^{-5}$ & $3.9\times 10^{-4}$ & $4.4\times 10^{-3}$ & $5.0\times 10^{-5}$ & $4.4\times 10^{-4}$ & $4.7\times 10^{-3}$ \\ \cline{2-11} 
& $\mathbf{\sigma}_{|| \Delta\theta ||}$ & $1.5\times 10^{-5}$ & $1.2\times 10^{-4}$ & $1.2 \times 10^{-3}$ & $1.1\times 10^{-5}$ & $8.9\times 10^{-5}$ & $9.2\times 10^{-4}$ & $1.3\times 10^{-5}$ & $1.1\times 10^{-4}$ & $1.1\times 10^{-3}$ \\ \hline\hline
\multirow{2}{*}{\textbf{14-Bus}} & $\mathbf{\mu}_{|| \Delta\theta ||}$ & $5.7 \times 10^{-5}$ & $5.0 \times 10^{-4}$ & $5.7\times 10^{-3}$ & $6.1\times 10^{-5}$ & $5.2\times 10^{-4}$ & $5.4\times 10^{-3}$ & $5.8\times 10^{-5}$ & $5.1\times 10^{-4}$ & $5.4\times 10^{-3}$ \\ \cline{2-11} 
 & $\mathbf{\sigma}_{|| \Delta\theta ||}$ & $1.3 \times 10^{-5}$ & $1.1\times 10^{-4}$ & $1.5\times 10^{-3}$ & $1.5\times 10^{-5}$ & $1.1\times 10^{-4}$ & $1.3\times 10^{-3}$ & $1.4\times 10^{-5}$ & $1.1\times 10^{-4}$ & $1.3\times 10^{-3}$ \\ \hline\hline
\multirow{2}{*}{\textbf{RTS}} & $\mathbf{\mu}_{|| \Delta\theta ||}$ & $1.3 \times 10^{-4}$ & $1.1 \times 10^{-3}$ & $1.0\times 10^{-2}$ & $8.0\times 10^{-5}$ & $7.7\times 10^{-4}$ & $8.0\times 10^{-3}$ & $1.2\times 10^{-4}$ & $1.1\times 10^{-3}$ & $1.0\times 10^{-2}$ \\ \cline{2-11} 
& $\mathbf{\sigma}_{|| \Delta\theta ||}$ & $3.6 \times 10^{-5}$ & $3.5\times 10^{-4}$ & $2.9\times 10^{-3}$ & $1.4\times 10^{-5}$ & $1.2\times 10^{-4}$ & $1.3\times 10^{-3}$ & $3.2\times 10^{-5}$ & $3.4\times 10^{-4}$ & $2.4\times 10^{-3}$ \\ \hline\hline
\multirow{2}{*}{\textbf{118-Bus}} & $\mathbf{\mu}_{|| \Delta\theta ||}$ & $1.2\times 10^{-4}$ & $1.0 \times 10^{-3}$ & $1.1\times 10^{-2}$ & $1.1 \times 10^{-4}$ & $8.9\times 10^{-4}$ & $9.7\times 10^{-3}$ & $1.2\times 10^{-4}$ & $1.0\times 10^{-3}$ & $1.1\times 10^{-2}$ \\ \cline{2-11} 
& $\mathbf{\sigma}_{|| \Delta\theta ||}$ & $2.3\times 10^{-5}$ & $2.0 \times 10^{-4}$ & $2.0\times 10^{-3}$ & $1.5\times 10^{-5}$ & $1.4\times 10^{-4}$ & $1.2\times 10^{-3}$ & $2.0\times 10^{-5}$ & $1.8\times 10^{-4}$ & $1.8\times 10^{-3}$ \\ \hline\hline
\multirow{2}{*}{\textbf{300-Bus}} & $\mathbf{\mu}_{|| \Delta\theta ||}$ & $1.2 \times 10^{-3}$ & $2.3 \times 10^{-3}$ & $1.9\times 10^{-2}$ & $1.5\times 10^{-4}$ & $1.2\times 10^{-3}$ & $1.1\times 10^{-2}$ & $1.1\times 10^{-3}$ & $1.9\times 10^{-3}$ & $1.8\times 10^{-2}$ \\ \cline{2-11} 
& $\mathbf{\sigma}_{|| \Delta\theta ||}$ & $1.6 \times 10^{-4}$ & $1.1\times 10^{-3}$ & $7.2\times 10^{-3}$ & $3.3\times 10^{-4}$ & $1.8\times 10^{-4}$ & $1.2\times 10^{-3}$ & $1.6\times 10^{-4}$ & $6.8\times 10^{-4}$ & $6.7\times 10^{-3}$ \\ \hline
\end{tabular}}
\end{table*}
\newpage
\subsection{Performance with Bad Data}

\begin{figure}[h]
    \centering
        \includegraphics[width=3.3in]{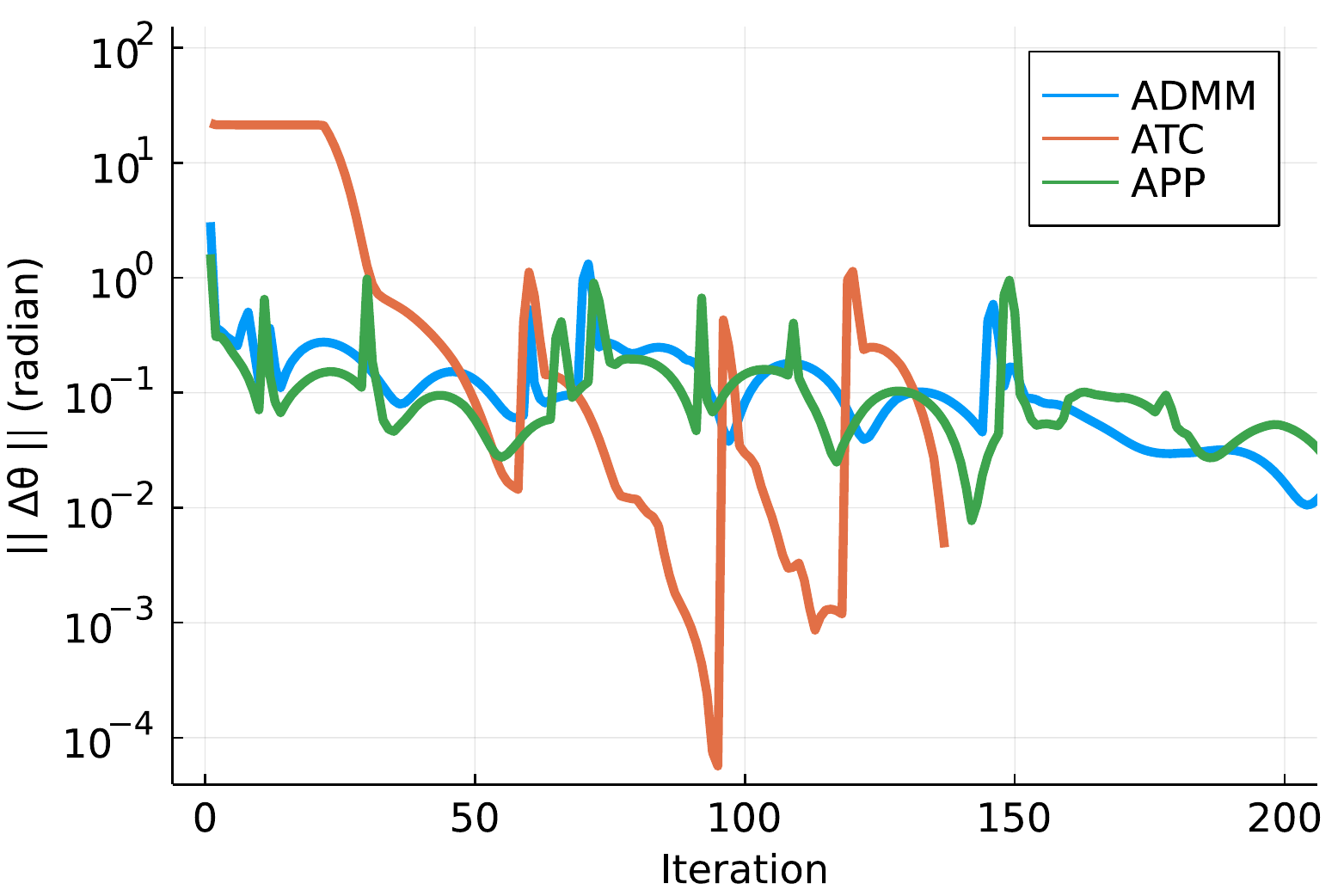}
	\caption[nt2convergence]{The three algorithms convergence for the IEEE 118-bus system with bad data ($p = 0.1\%$).}
	\label{fig:nt2_1}
\end{figure}

For the second type of noise, we inject bad data into the shared variables as described in~\eqref{nt:2}. We set the value for the bad data magnitude $R = 2$ radians and vary the probability of the injected errors. Fig.~\ref{fig:nt2_1} shows the mismatches of the shared variables with probabilities of bad data occurrence $p =0.1\%$ for the IEEE 118-bus system. 

\begin{figure}[h!]
    \centering
        \includegraphics[width=3.2in]{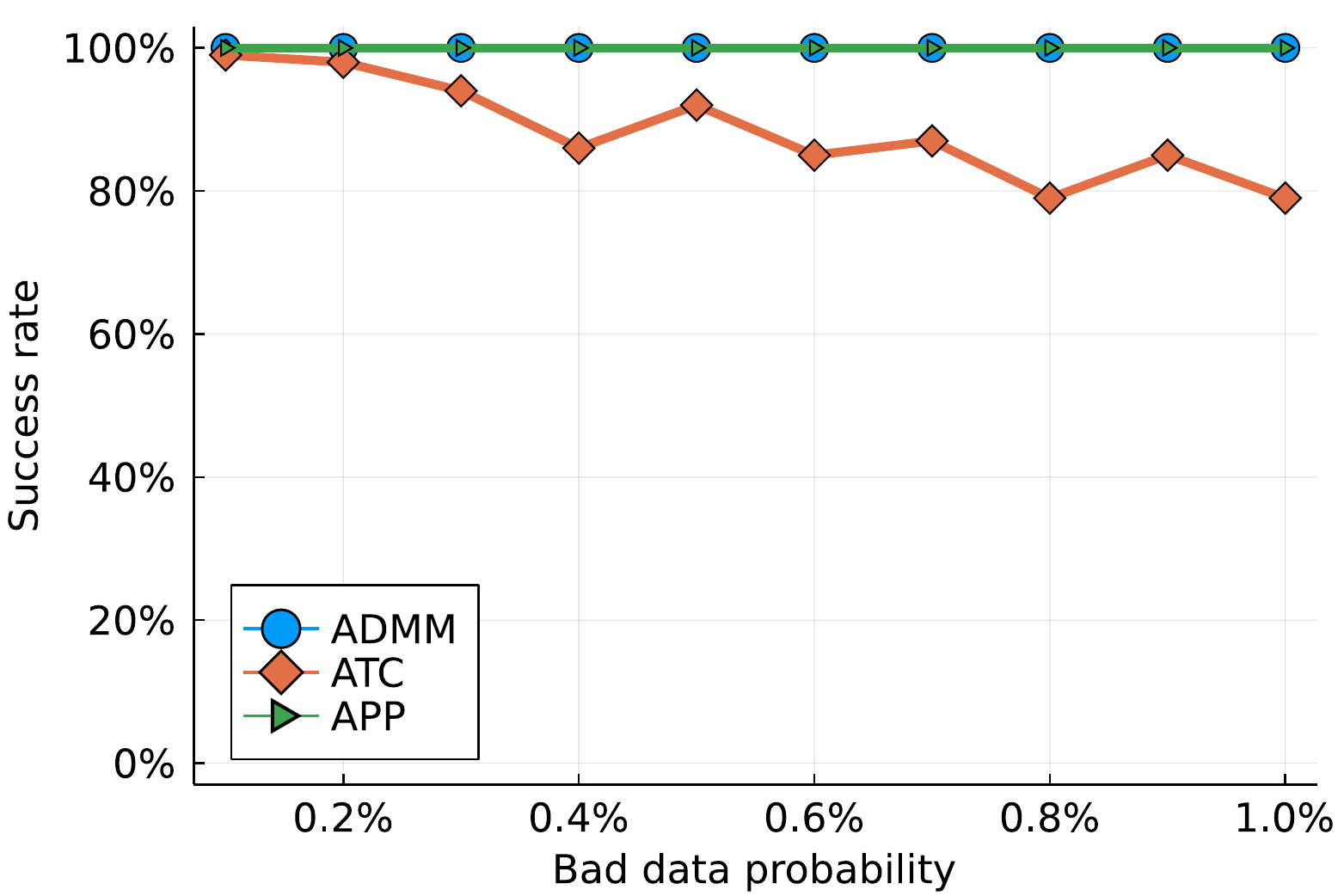}
	\caption[nt2comparison]{Success rates for the WB5 system with different rates of bad data injection.}
	\label{fig:nt2_comparison_5}
\end{figure}

\begin{figure}[h!]
    \centering
        \includegraphics[width=3.2in]{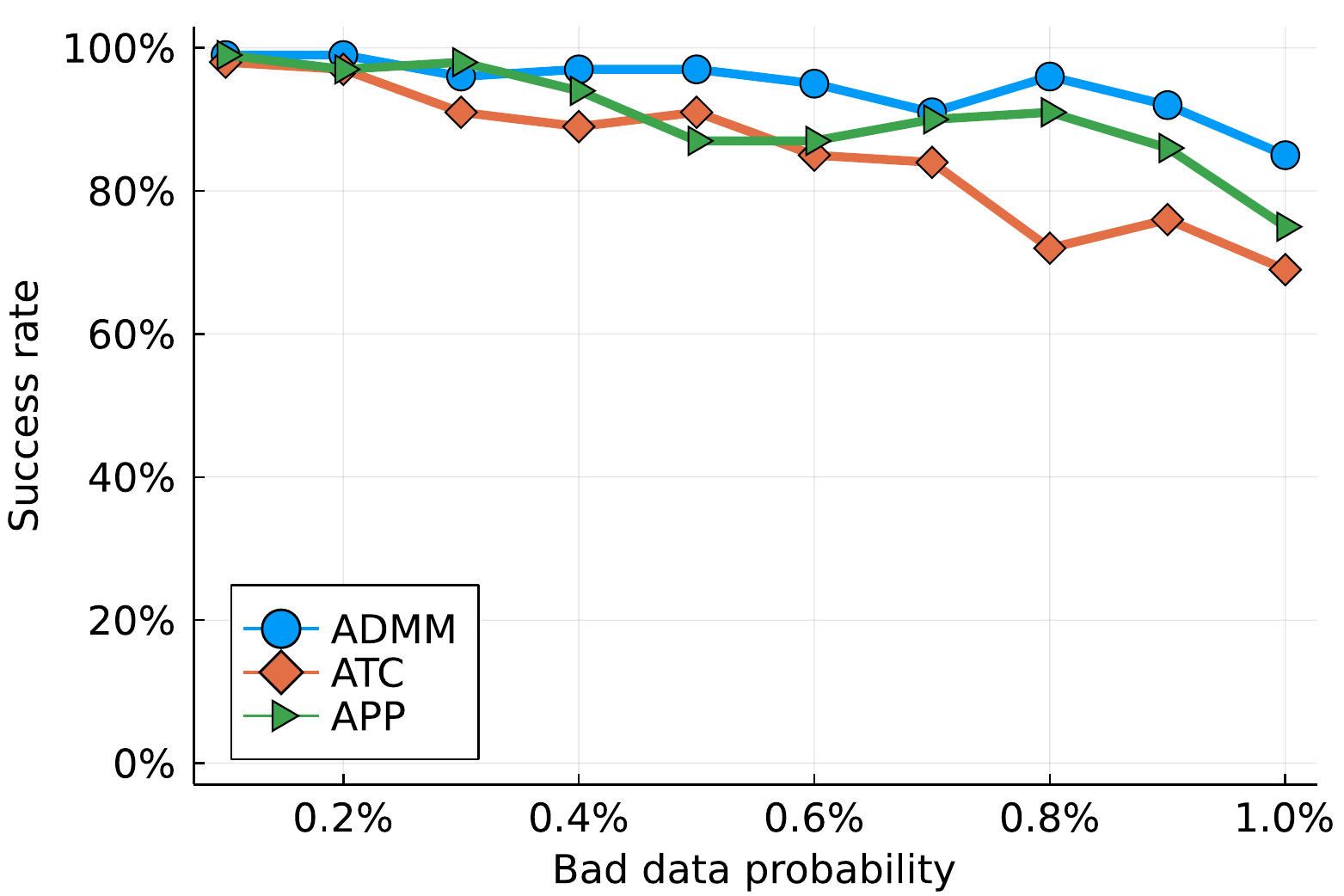}
	\caption[nt2comparison]{Success rates for the IEEE 14-bus system with different rates of bad data injection.}
	\label{fig:nt2_comparison_14}
\end{figure}

\begin{figure}[h!]
    \centering
        \includegraphics[width=3.2in]{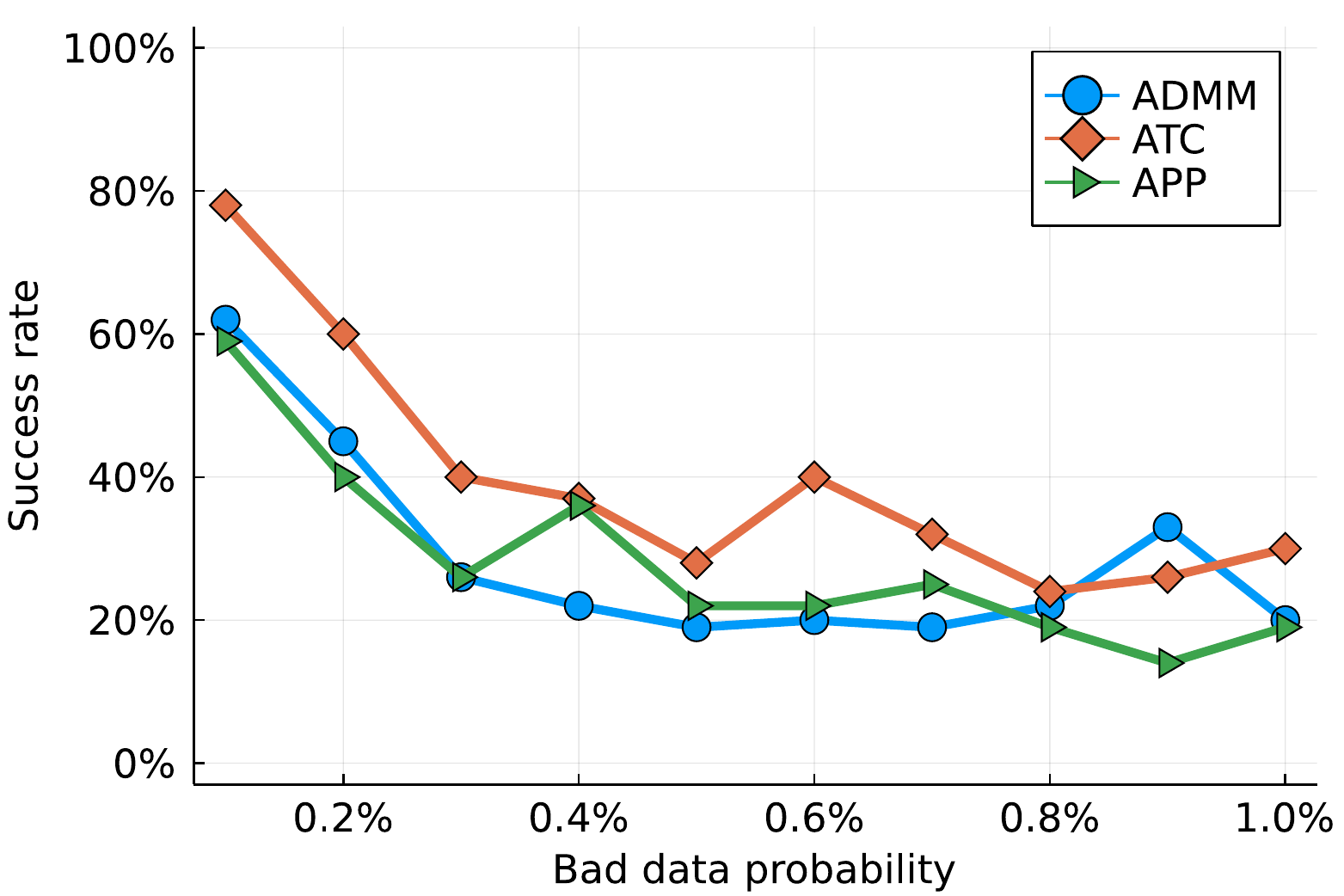}
	\caption[nt2comparison]{Success rates for the RTS GMLC system with different rates of bad data injection.}
	\label{fig:nt2_comparison_rts}
\end{figure}

\begin{figure}[h!]
    \centering
        \includegraphics[width=3.2in]{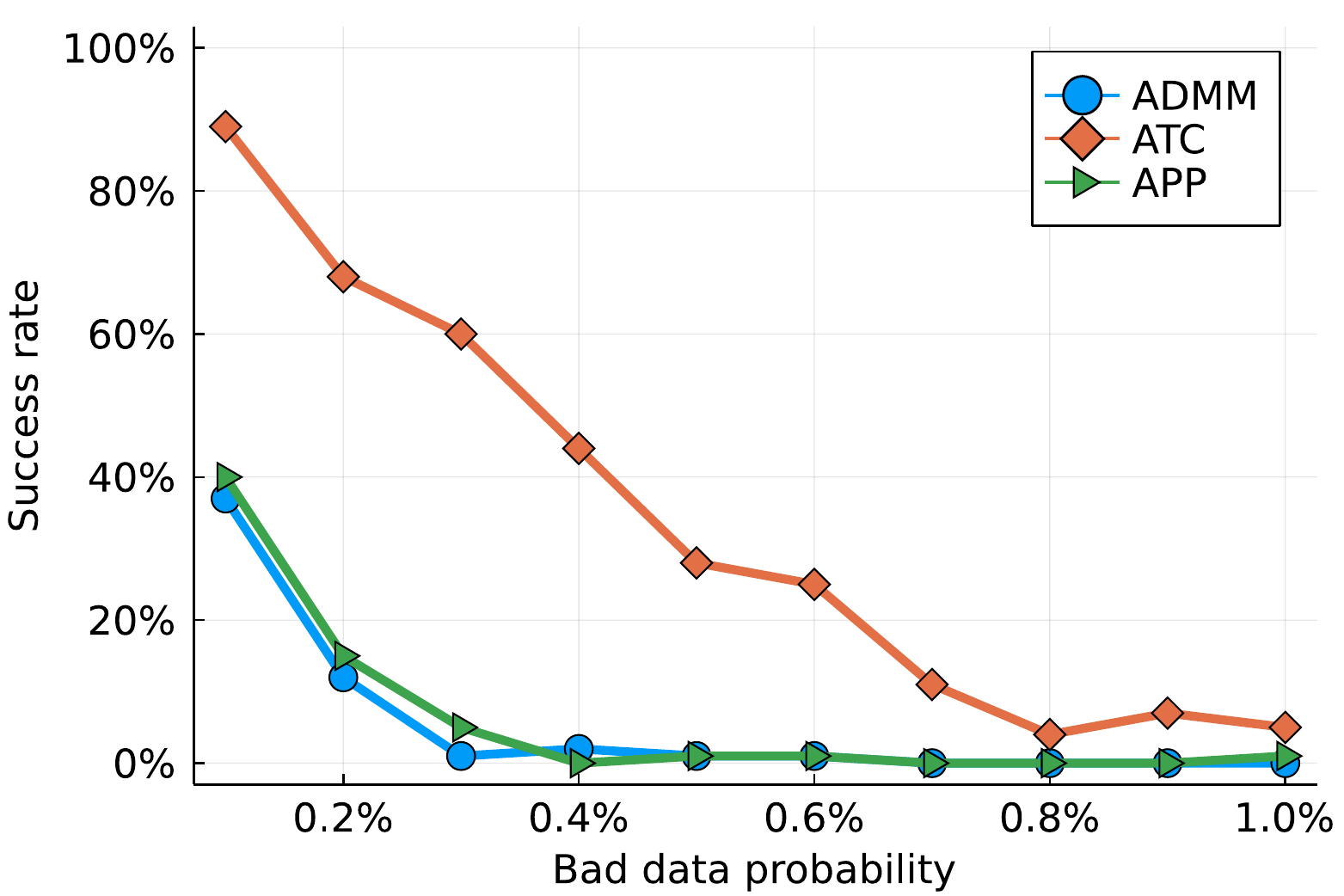}
	\caption[nt2comparison]{Success rates for the IEEE 118-bus system with different rates of bad data injection.}
	\label{fig:nt2_comparison_118}
\end{figure}

\begin{figure}[h!]
    \centering
        \includegraphics[width=3.2in]{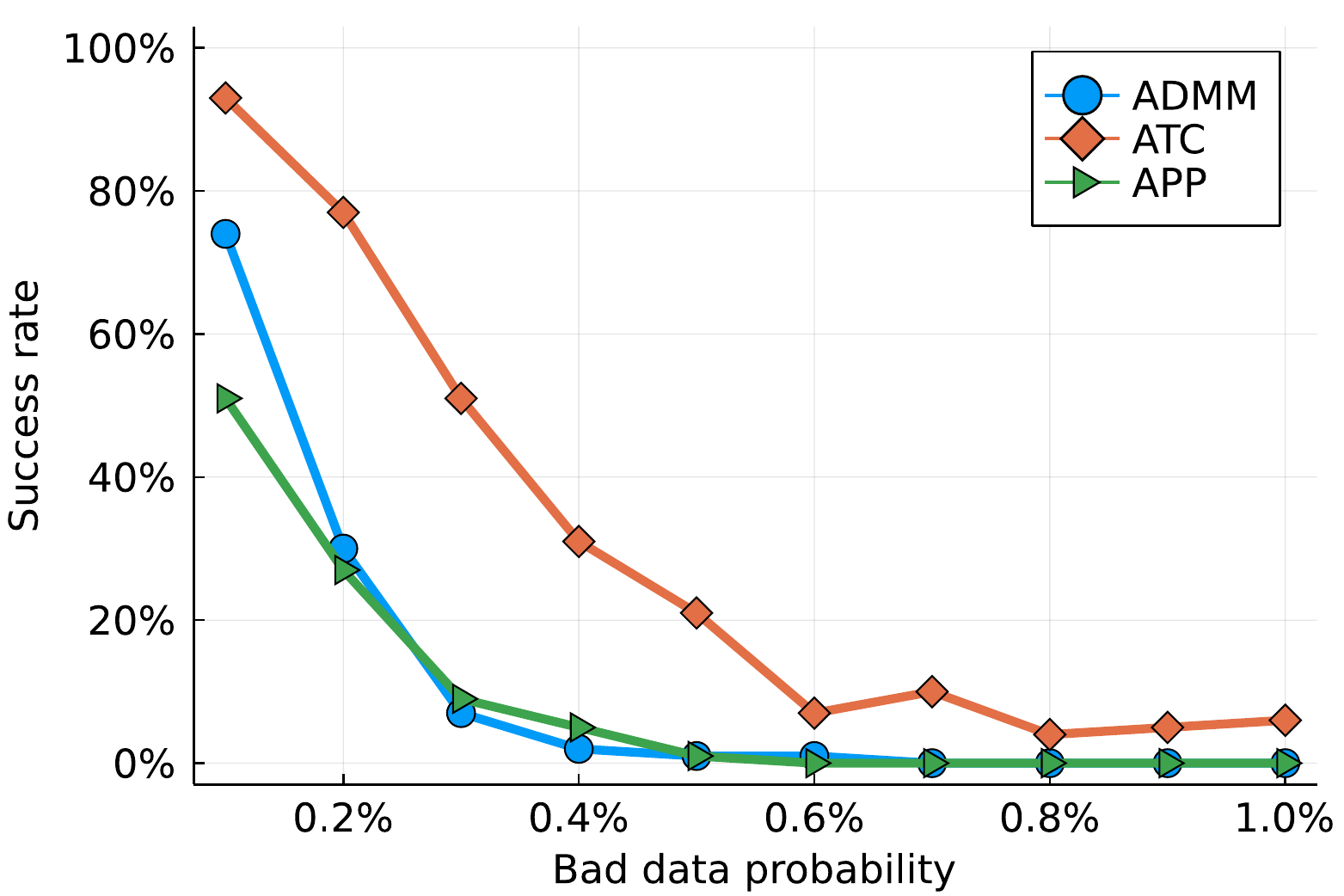}
	\caption[nt2comparison]{Success rates for the IEEE 300-bus system with different rates of bad data injection.}
	\label{fig:nt2_comparison_300}
\end{figure}

We observe that the mismatches return to the same convergence pattern after a large error is injected in all three algorithms. To quantitatively compare the algorithms' performance, we estimate the probability of achieving the optimal solution using 100 runs of the algorithm for varying probabilities of bad data occurrence. We consider an algorithm to have achieved the optimal solution if the value of the relative gap is below $1\%$ within a maximum of 1000 iterations. Table~\ref{Tb3} on the next page summarizes the results with probabilities of bad data equal to $0.1\%$ and $1\%$. The results show that the distributed algorithms are highly susceptibility to the bad data and they may fail to attain the optimal solution even when the bad data probability is as low as $0.1\%$. Further, Fig.~\ref{fig:nt2_comparison_5}-\ref{fig:nt2_comparison_300} show the success rates when varying the bad data probability from $0.1\%$ to $1\%$ for the test systems. Comparing the three algorithms, the ATC algorithm shows a better performance to bad data errors. For the IEEE 118-bus, IEEE 300-bus, and RTS~GMLC test systems, the ATC algorithm attains the optimal solution more than 80\% of the time when the probability of the bad data is around 0.1\%, while the ADMM and APP fail 40\% of the time in the same cases.

\begin{table*}[t]
\centering
\caption{Performance of the distributed algorithms with bad data}
\label{Tb3}
\begin{tabular}{|l|l||r|r||r|r||r|r|}
\hline
\multicolumn{2}{|c||}{\textbf{Algorithm}} & \multicolumn{2}{c||}{\textbf{ADMM}} & \multicolumn{2}{c||}{\textbf{ATC}} & \multicolumn{2}{c|}{\textbf{APP}} \\ \hline
\multicolumn{2}{|c||}{\textbf{Probability of Bad Data}} & \multicolumn{1}{c|}{$\mathbf{0.1\%}$} & \multicolumn{1}{c||}{$\mathbf{1.0\%}$} & \multicolumn{1}{c|}{$\mathbf{0.1\%}$} & \multicolumn{1}{c||}{$\mathbf{1.0\%}$} & \multicolumn{1}{c|}{$\mathbf{0.1\%}$} & \multicolumn{1}{c|}{$\mathbf{1.0\%}$} \\ \hline\hline
\multirow{2}{*}{\textbf{WB5}} & \textbf{Success rate (\%)} & 100 & 100 & 100 & 100 & 100 & 76 \\ \cline{2-8} 
 & \textbf{Avg. Iterations} & 32 & 41 & 16 & 17 & 25 & 31 \\ \hline\hline
\multirow{2}{*}{\textbf{14-Bus}} & \textbf{Success rate (\%)} & 100 & 85 & 99 & 70 & 100 & 88 \\ \cline{2-8} 
 & \textbf{Avg. Iterations} & 32 & 59 & 44 & 48 & 21 & 43 \\ \hline\hline
\multirow{2}{*}{\textbf{RTS}} & \textbf{Success rate (\%)} & 72 & 19 & 83 & 29 & 56 & 19 \\ \cline{2-8} 
 & \textbf{Avg. Iterations} & 96 & 1000 & 66 & 109 & 104 & 1000 \\  \hline\hline
\multirow{2}{*}{\textbf{118-Bus}} & \textbf{Success rate (\%)} & 30 & 0 & 83 & 8 & 30 & 0 \\ \cline{2-8} 
 & \textbf{Avg. Iterations} & 123 & NC & 60 & 66 & 95 & NC \\  \hline\hline
\multirow{2}{*}{\textbf{300-Bus}} & \textbf{Success rate (\%)} & 62 & 0 & 92 & 3 & 55 & 0 \\ \cline{2-8} 
 & \textbf{Avg. Iterations} & 34 & NC & 62 & 65 & 34 & NC \\  \hline
\end{tabular}

\vspace{4pt}

{\raggedright \qquad\qquad\qquad\qquad\qquad\qquad\qquad\quad \ NC: Not converged. \par}
\end{table*}

\newpage
\subsection{Performance with Intermittent Communication Loss}

\begin{figure}[h]
    \centering
        \includegraphics[ width=3.3in]{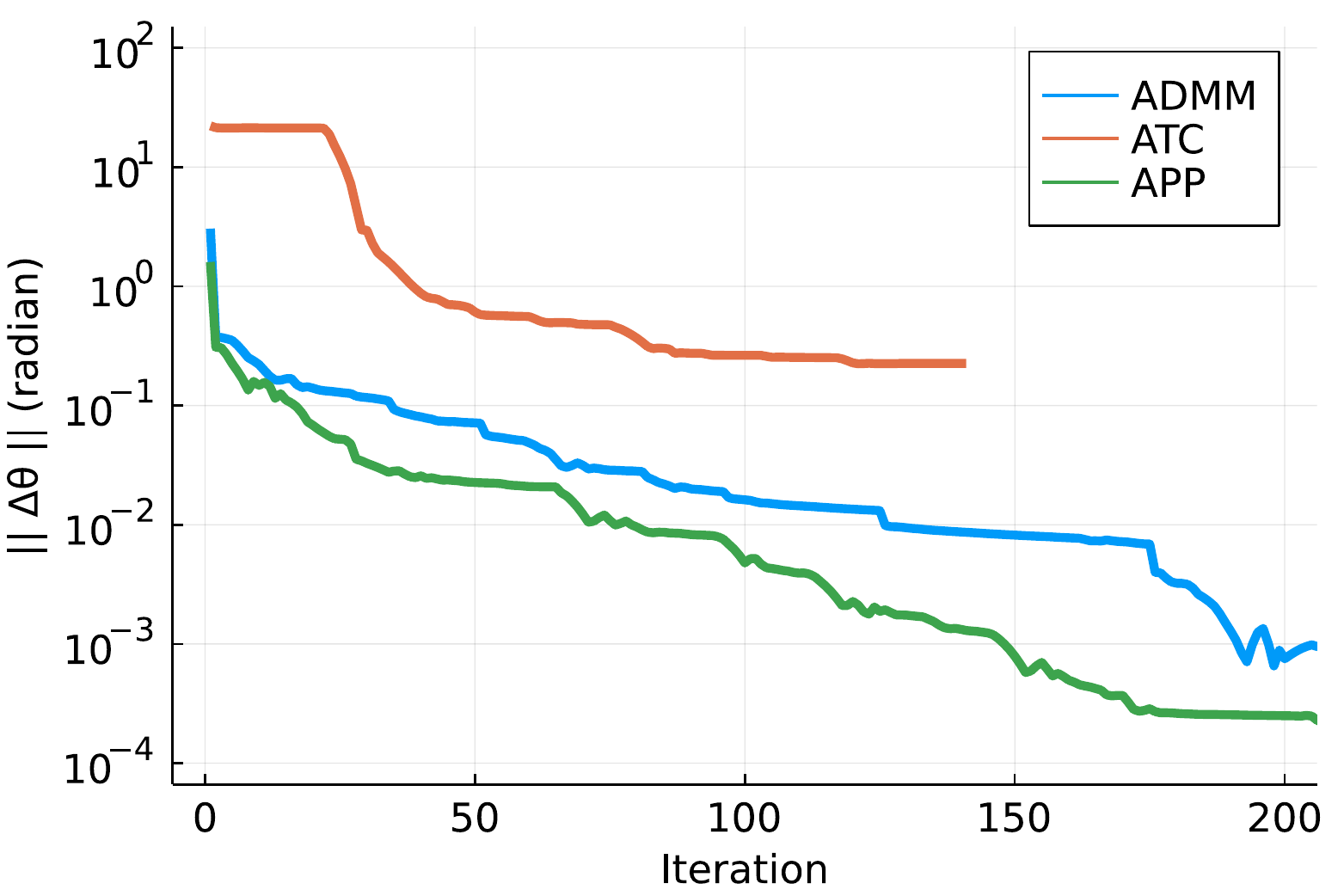}
	\caption[nt3convergence]{The three algorithms convergence for the IEEE 118-bus system with intermittent communication loss ($\lambda_f = 10\%$, $\lambda_r = 10\%$).}
	\label{fig:nt3_1}
\end{figure}

This section compares the distributed algorithms' performance under the intermittent communication loss model in Section~\ref{nt3}. Fig.~\ref{fig:nt3_1} shows the convergence characteristics of the shared variable mismatches for the three algorithms with failure probability $\lambda_f = 10\%$ and repair probability $\lambda_r = 10\%$.

The results indicate that the distributed algorithms achieve consensus on the values of the shared variables and the mismatch almost always decreases as the number of iterations increases for all cases. However, the final solutions can be far from optimal with high values of the relative gap. To qualitatively compare algorithmic performance during intermittent communication loss, Table~\ref{Tb4} describes how the probability of achieving the optimal solution changes with failure probability equal to $1\%$ and $5\%$ per iteration. Each of the results in this table are computed from $100$ runs of the algorithm with a constant repair rate of $10\%$ per iteration. We again consider an algorithm to have achieved the optimal solution if the relative gap is below $1\%$ within a maximum of $1000$ iterations.

The results show that the three algorithm performance is impacted by the communication loss values to varying degrees. With a low probability of failure, the algorithms manage to archive the optimal solution most of the time. Comparing the three algorithms, the ATC algorithm is the most susceptible algorithm to communication loss. The ATC algorithm fails around half the time to obtain the optimal solution for the largest three test system when the failure probability is $1\%$, and almost all the time when the failure probability is $5\%$.

\begin{figure}[h]
    \centering
        \includegraphics[  width=3.3in]{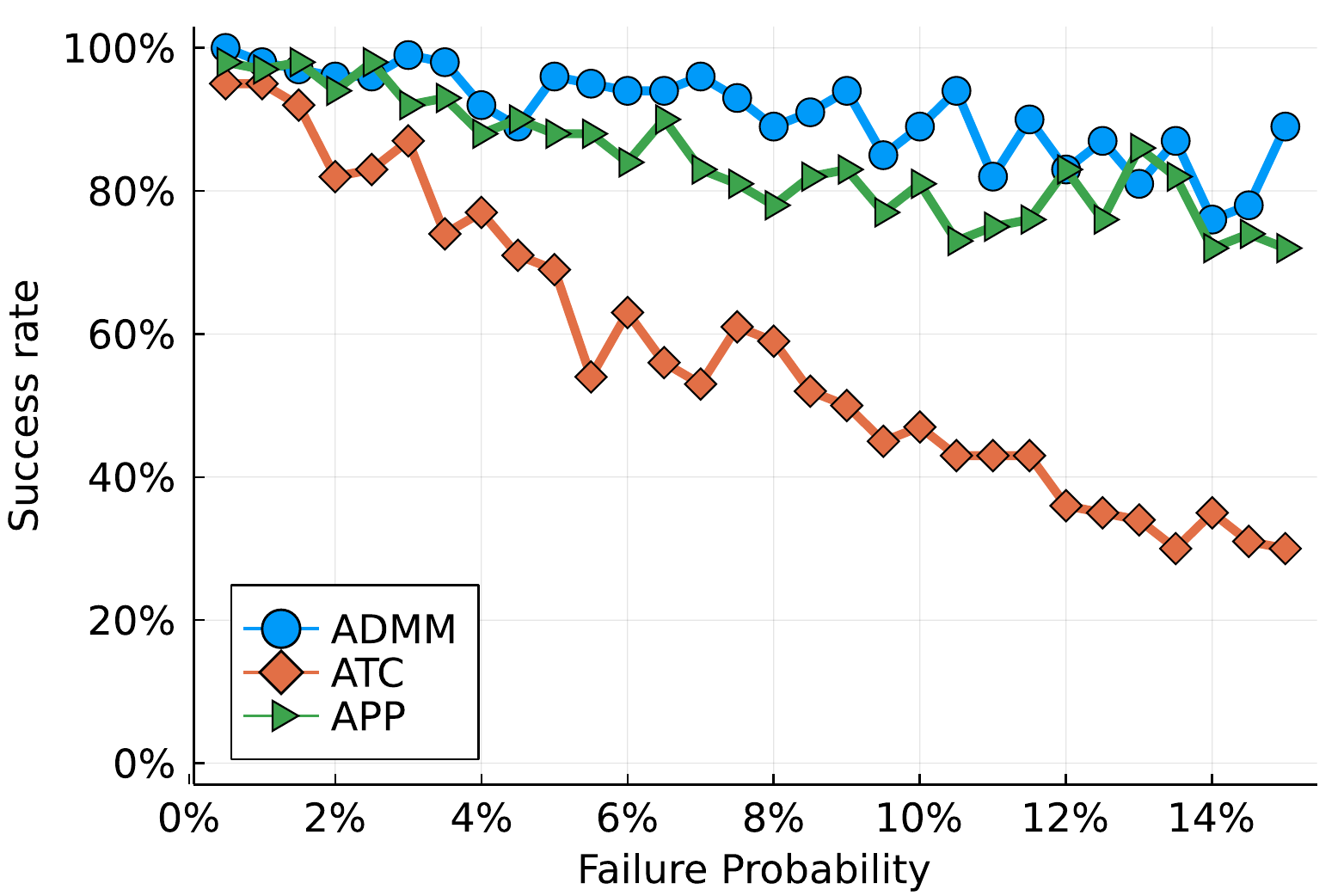}
	\caption[nt3comparison]{Success rates for the WB5 system with different communication failure probability. Communication repair probability $\lambda_r = 10\%$.}
	\label{fig:nt3_comparison_5}
\end{figure}

\begin{figure}[h]
    \centering
        \includegraphics[  width=3.3in]{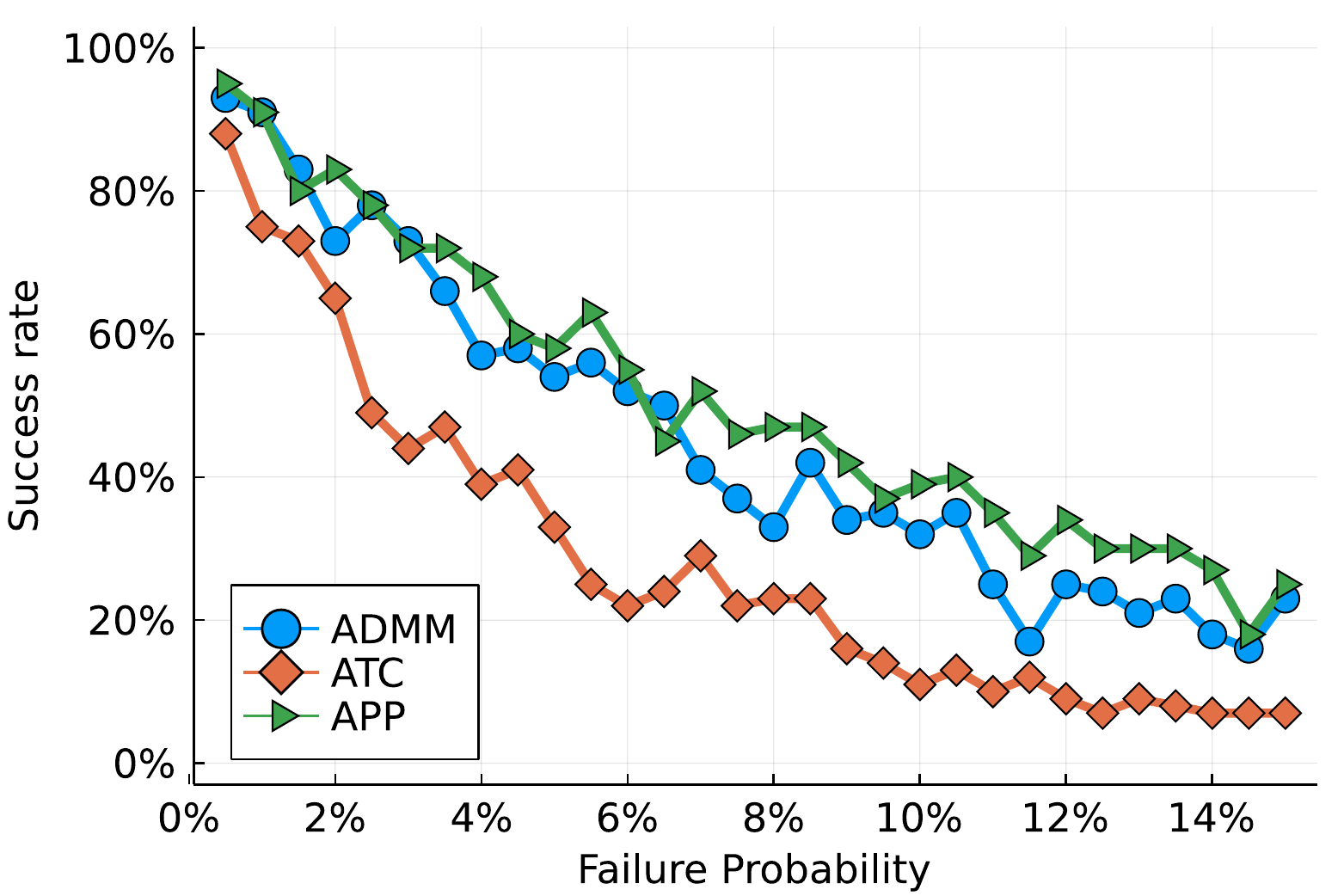}
	\caption[nt3comparison]{Success rates for the IEEE 14-bus system with different communication failure probability. Communication repair probability $\lambda_r = 10\%$.}
	\label{fig:nt3_comparison_14}
\end{figure}

\begin{figure}[h]
    \centering
        \includegraphics[  width=3.3in]{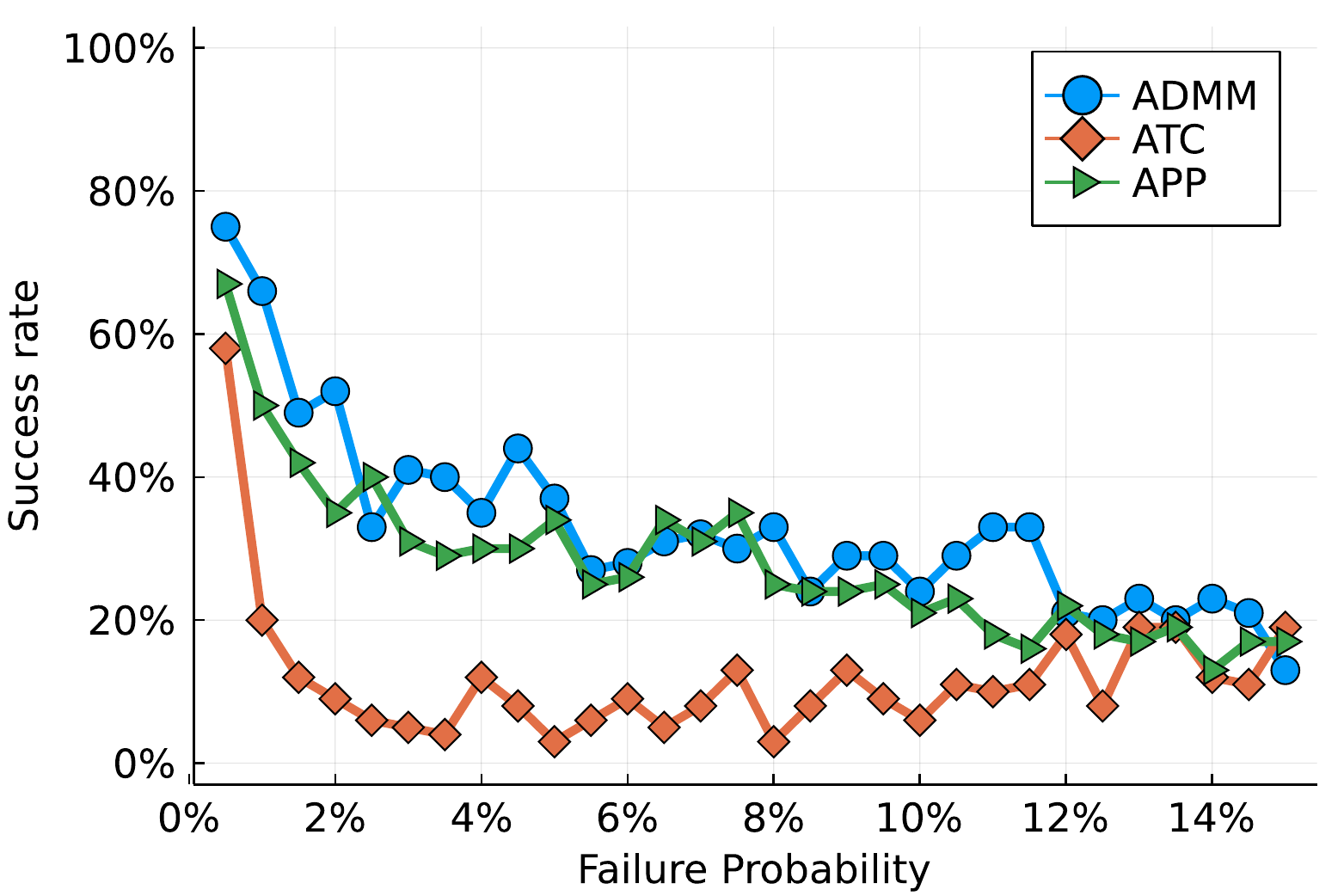}
	\caption[nt3comparison]{Success rates for the RTS GMLC system with different communication failure probability. Communication repair probability $\lambda_r = 10\%$.}
	\label{fig:nt3_comparison_rts}
\end{figure}

\begin{figure}[h]
    \centering
        \includegraphics[  width=3.3in]{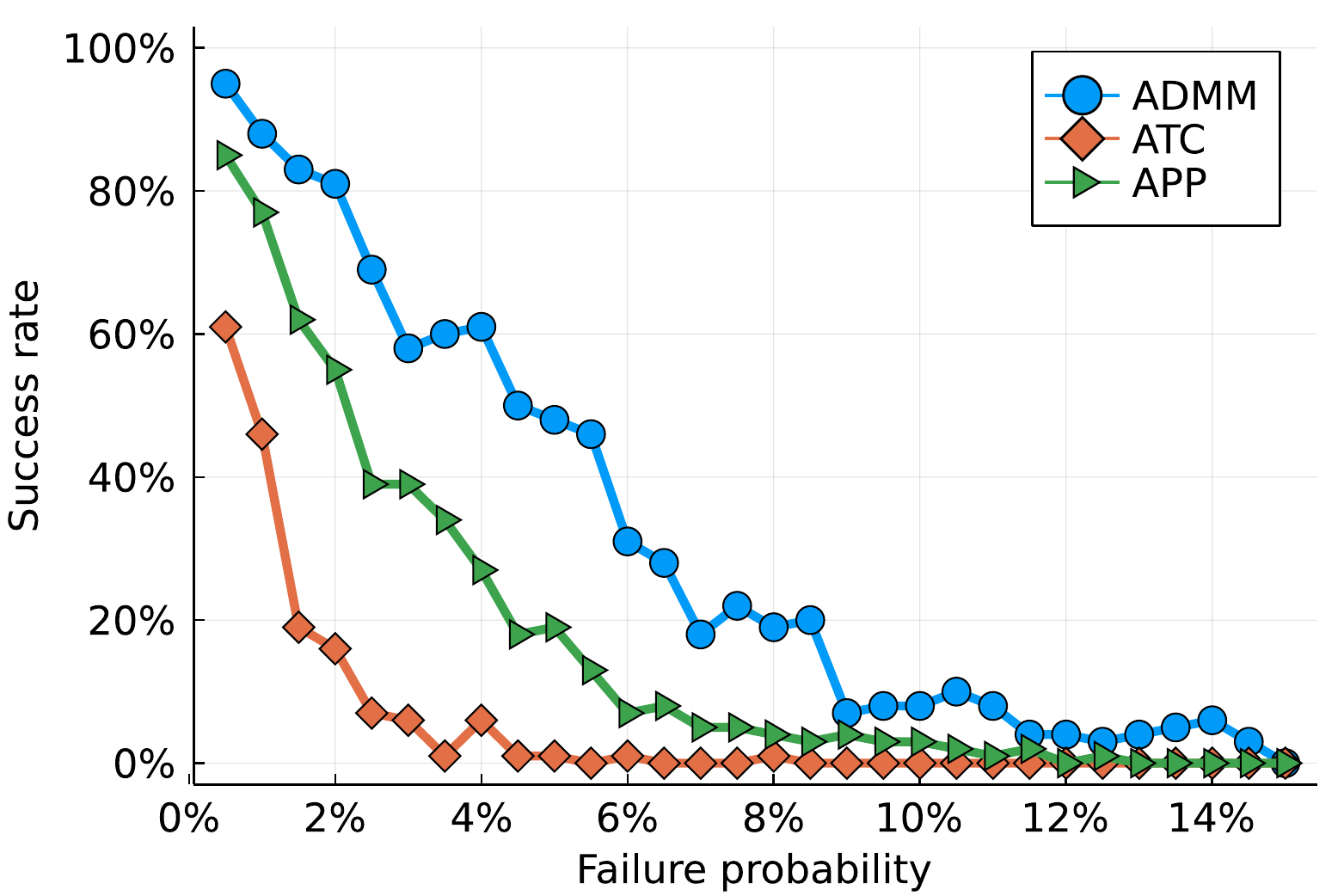}
	\caption[nt3comparison]{Success rates for the IEEE 118-bus system with different communication failure probability. Communication repair probability $\lambda_r = 10\%$.}
	\label{fig:nt3_comparison_118}
\end{figure}

\begin{figure}[h!]
    \centering
        \includegraphics[  width=3.3in]{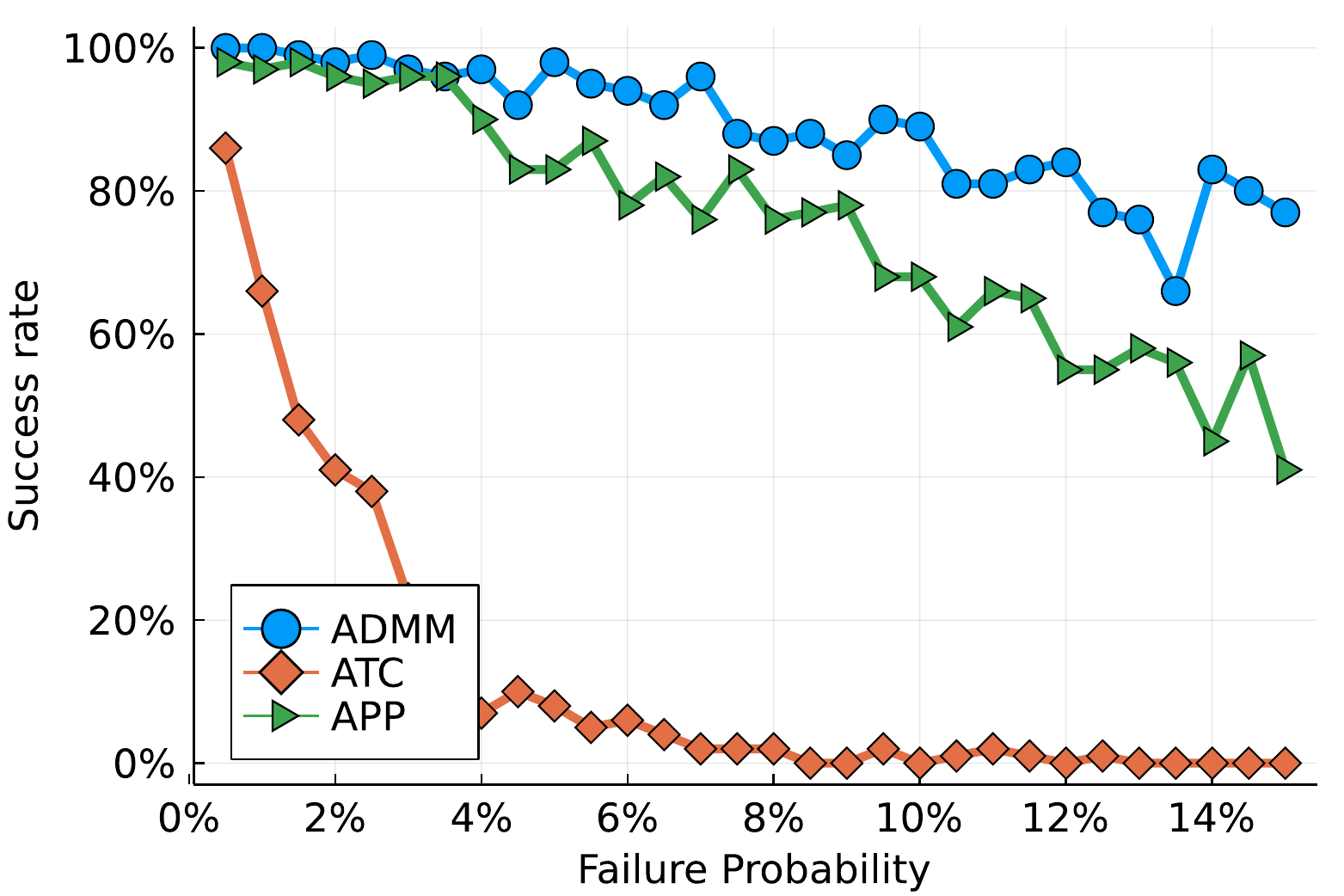}
	\caption[nt3comparison]{Success rates for the IEEE 300-bus system with different communication failure probability. Communication repair probability $\lambda_r = 10\%$.}
	\label{fig:nt3_comparison_300}
\end{figure}

Fig.~\ref{fig:nt3_comparison_5}-~\ref{fig:nt3_comparison_300} present a comparison between the algorithms' performance when subjected to intermittent communication loss by varying the failure probability from $0.5\%$ to $15\%$, while fixing the repair probability to $10\%$. The ADMM algorithm shows a slightly better performance than the APP algorithm for the five test systems. Overall, the results suggest that the ADMM and APP algorithms outperform the ATC algorithm with the intermittent communication loss model for all test systems we considered in this study.

\begin{table*}[t]
\centering
\caption{Performance of the distributed algorithms with Intermittent Communication Loss}
\label{Tb4}
\begin{tabular}{|l|l||r|r||r|r||r|r|}
\hline
\multicolumn{2}{|c||}{\textbf{Algorithm}} &
\multicolumn{2}{c||}{\textbf{ADMM}} &
\multicolumn{2}{c||}{\textbf{ATC}} &
\multicolumn{2}{c|}{\textbf{APP}} \\ \hline
\multicolumn{2}{|c||}{\textbf{Failure Probability}} &
\multicolumn{1}{c|}{$\mathbf{1\%}$} &
\multicolumn{1}{c||}{$\mathbf{5\%}$} &
\multicolumn{1}{c|}{$\mathbf{1\%}$} &
\multicolumn{1}{c||}{$\mathbf{5\%}$} &
\multicolumn{1}{c|}{$\mathbf{1\%}$} &
\multicolumn{1}{c|}{$\mathbf{5\%}$} \\ \hline\hline
\multirow{2}{*}{\textbf{WB5}} & \textbf{Success rate (\%)} & 97 & 88 & 92 & 71 & 99 & 89 \\ \cline{2-8} 
 & \textbf{Avg. Iterations} & 36 & 45 & 17 & 24 & 27 & 37 \\  \hline\hline
\multirow{2}{*}{\textbf{14-Bus}} & \textbf{Success rate (\%)} & 84 & 69 & 82 & 38 & 94 & 57 \\ \cline{2-8} 
 & \textbf{Avg. Iterations} & 32 & 34 & 43 & 45 & 22 & 26 \\  \hline\hline
\multirow{2}{*}{\textbf{RTS}} & \textbf{Success rate} (\%) & 70 & 28 & 33 & 7 & 58 & 29 \\ \cline{2-8} 
 & \textbf{Avg. Iterations} & 81 & 91 & 69 & 124 & 71 & 94 \\  \hline\hline
\multirow{2}{*}{\textbf{118-Bus}} & \textbf{Success rate} (\%) & 88 & 48 & 46 & 1 & 77 & 19 \\ \cline{2-8} 
 & \textbf{Avg. Iterations} & 89 & 77 & 60 & 54 & 75 & 61 \\  \hline\hline
\multirow{2}{*}{\textbf{300-Bus}} & \textbf{Success rate} (\%) & 99 & 95 & 65 & 6 & 99 & 89 \\ \cline{2-8} 
 & \textbf{Avg. Iterations} & 37 & 71 & 65 & 75 & 35 & 71 \\  \hline
\end{tabular}
\end{table*}

\subsection{Discussion and Comparison}
Parameter tuning plays a major role in the performance of distributed algorithms as it strongly impacts the convergence rate. The parameters in the selected algorithms are associated with the penalty terms for the relaxed consistency constraints. We observe that all three algorithms converge for a certain range of parameter values. Generally speaking, selecting large parameter values will prevent the algorithm from achieving the optimal solution and, in some cases, large values might cause the algorithm to diverge. On the other hand, small values reduce the convergence rate and, in extreme cases, the algorithm can diverge. We also observe that different systems and cost functions might require repeating the parameter tuning step.

The results shown in this paper indicate the importance of data integrity on the performance of the distributed algorithms. We observe various responses from the distributed algorithms to the error models. With Gaussian communication noise, all three algorithms converge to an accuracy that is proportional to the standard deviation of the noise. We observed a slightly better performance when using the ATC algorithm compared to the other two algorithms. Among the three algorithms considered in this paper, the ATC algorithm has the best performance with the presence of bad data, while the ADMM and APP algorithms have the best performance when there is a high intermittent communication loss probability.

Both the ADMM and APP algorithms have a very similar performance pattern for the three noise models with slightly better performance observed when using the ADMM algorithm. The ATC algorithm on the other hand has a different performance pattern. Further, the ATC algorithm's final solution can have a high relative gap from the optimal solution or lead to numerical instability in the optimization solver if consensus is not achieved, i.e., the stopping criteria are not met after many iterations. This happens due to the parameter update for the ATC algorithm~\eqref{eq:ATC4}, which exponentially increases the penalty on the consistency term in the objective as the number of iterations increases. This suggests that reliable performance of the ATC algorithm in the presence of noise requires either using another stopping criterion or modifying the update step in the algorithm.

\section{Conclusion and Future Work} \label{SEC6}

Distributed algorithms have many attractive features for solving power system optimization problems, especially for systems with many independent microgrids. Distributed algorithms allow interconnected systems to cooperatively solve large optimization problems while maintaining their autonomy. However, the performance of a distributed algorithm strongly depends on the quality of the shared data. In this paper, we numerically show distributed algorithms' responses to data quality issues. We evaluate the performance of ADMM, ATC, and APP distributed algorithms using three noise models.

The results show that the three algorithms perform well with additive Gaussian noise as long as the stopping criteria and the required solution accuracy are lower than the error standard deviation. The results also show that the bad data errors have a severe impact on the quality of the distributed algorithms' solutions even with low error probability. Moreover, the impacts of intermittent communication loss might not be visible by the local controllers, as the distributed algorithms might reach a consensus on the shared variables corresponding to an operating point that is far from optimal.

As extensions to this work, there are other communication and data integrity issues requiring detailed investigations. For power systems applications, these include asynchronous data sharing between neighboring regions and communication latency. We further plan to use hardware-in-the-loop testing with an actual communication network to investigate the performance of distributed algorithms in practical setups. Another direction for future work is studying how different power flow representations affect the convergence rates for problems where the DC approximation is inapplicable.

\ifCLASSOPTIONcaptionsoff
  \newpage
\fi



%
\bibliographystyle{IEEEtran}
\bibliography{IEEEabrv,manuscript}

%








\end{document}